\PassOptionsToPackage{unicode}{hyperref}
\PassOptionsToPackage{hyphens}{url}
\PassOptionsToPackage{dvipsnames,svgnames,x11names}{xcolor}
\documentclass[
  letterpaper,
  DIV=11,
  numbers=noendperiod]{scrartcl}

\usepackage{amsmath,amssymb}
\usepackage{iftex}
\ifPDFTeX
  \usepackage[T1]{fontenc}
  \usepackage[utf8]{inputenc}
  \usepackage{textcomp} 
\else 
  \usepackage{unicode-math}
  \defaultfontfeatures{Scale=MatchLowercase}
  \defaultfontfeatures[\rmfamily]{Ligatures=TeX,Scale=1}
\fi
\usepackage{lmodern}
\ifPDFTeX\else  
\fi
\IfFileExists{upquote.sty}{\usepackage{upquote}}{}
\IfFileExists{microtype.sty}{
  \usepackage[]{microtype}
  \UseMicrotypeSet[protrusion]{basicmath} 
}{}
\makeatletter
\@ifundefined{KOMAClassName}{
  \IfFileExists{parskip.sty}{%
    \usepackage{parskip}
  }{
    \setlength{\parindent}{0pt}
    \setlength{\parskip}{6pt plus 2pt minus 1pt}}
}{
  \KOMAoptions{parskip=half}}
\makeatother
\usepackage{xcolor}
\ifx\paragraph\undefined\else
  \let\oldparagraph\paragraph
  \renewcommand{\paragraph}[1]{\oldparagraph{#1}\mbox{}}
\fi
\ifx\subparagraph\undefined\else
  \let\oldsubparagraph\subparagraph
  \renewcommand{\subparagraph}[1]{\oldsubparagraph{#1}\mbox{}}
\fi

\usepackage{longtable,booktabs,array}
\usepackage{calc} 
\usepackage{etoolbox}
\makeatletter
\patchcmd\longtable{\par}{\if@noskipsec\mbox{}\fi\par}{}{}
\makeatother
\IfFileExists{footnotehyper.sty}{\usepackage{footnotehyper}}{\usepackage{footnote}}
\makesavenoteenv{longtable}
\usepackage{graphicx}
\makeatletter
\def\maxwidth{\ifdim\Gin@nat@width>\linewidth\linewidth\else\Gin@nat@width\fi}
\def\maxheight{\ifdim\Gin@nat@height>\textheight\textheight\else\Gin@nat@height\fi}
\makeatother
\setkeys{Gin}{width=\maxwidth,height=\maxheight,keepaspectratio}
\makeatletter
\def\fps@figure{htbp}
\makeatother
\NewDocumentCommand\citeproctext{}{}

\makeatletter
 \let\@cite@ofmt\@firstofone
 \def\@biblabel#1{}
 \def\@cite#1#2{{#1\if@tempswa , #2\fi}}
\makeatother
\newlength{\cslhangindent}
\newlength{\csllabelwidth}
\newenvironment{CSLReferences}[2] 
 {\begin{list}{}{%
  \setlength{\itemindent}{0pt}
  \setlength{\leftmargin}{0pt}
  \setlength{\parsep}{0pt}
  \ifodd #1
   \setlength{\leftmargin}{\cslhangindent}
   \setlength{\itemindent}{-1\cslhangindent}
  \fi
  \setlength{\itemsep}{#2\baselineskip}}}
 {\end{list}}
\usepackage{calc}

\KOMAoption{captions}{tableheading}
\makeatletter
\@ifpackageloaded{caption}{}{\usepackage{caption}}
\AtBeginDocument{%
\ifdefined\contentsname
  \renewcommand*\contentsname{Table of contents}
\else
  \newcommand\contentsname{Table of contents}
\fi
\ifdefined\listfigurename
  \renewcommand*\listfigurename{List of Figures}
\else
  \newcommand\listfigurename{List of Figures}
\fi
\ifdefined\listtablename
  \renewcommand*\listtablename{List of Tables}
\else
  \newcommand\listtablename{List of Tables}
\fi
\ifdefined\figurename
  \renewcommand*\figurename{Figure}
\else
  \newcommand\figurename{Figure}
\fi
\ifdefined\tablename
  \renewcommand*\tablename{Table}
\else
  \newcommand\tablename{Table}
\fi
}
\@ifpackageloaded{float}{}{\usepackage{float}}
\floatstyle{ruled}
\@ifundefined{c@chapter}{\newfloat{codelisting}{h}{lop}}{\newfloat{codelisting}{h}{lop}[chapter]}
\floatname{codelisting}{Listing}

\makeatother
\makeatletter
\@ifpackageloaded{caption}{}{\usepackage{caption}}
\@ifpackageloaded{subcaption}{}{\usepackage{subcaption}}
\makeatother
\ifLuaTeX
  \usepackage{selnolig}  
\fi
\usepackage{bookmark}

\IfFileExists{xurl.sty}{\usepackage{xurl}}{} 
\hypersetup{
  pdftitle={Gender gap in the desired wages: Evidence from large administrative data},
  pdfauthor={Taiyo Fukai; Keisuke Kawata; Mizuki Komura; Takahiro Toriyabe},
  colorlinks=true,
  linkcolor={blue},
  filecolor={Maroon},
  citecolor={Blue},
  urlcolor={Blue},
  pdfcreator={LaTeX via pandoc}}

\usepackage{setspace}

\title{Gender gap in the desired wages:\\ Evidence from large administrative data}
\author{Taiyo Fukai\footnote{Faculty of Economics, Gakushuin University,
  1-5-1 Mejiro, Toshima-ku, Tokyo 171-8588, Japan,
  Email:taiyo.fukai@gakushuin.ac.jp} \and Keisuke
Kawata\footnote{Institute of Social Sciences, University of Tokyo, 7-3-1
  Hongo, Bunkyo-ku, Tokyo 113-0033, Japan,
  Email:keisukekawata@iss.u-tokyo.ac.jp} \and Mizuki
Komura\footnote{School of Economics, Kwansei Gakuin University, 1-155,
  Uegahara Ichibancho, Nishinomiya 662-8501, Japan,
  Email:m.komura@kwansei.ac.jp} \and Takahiro Toriyabe\footnote{Graduate
  School of Economics, Hitotsubashi University, 2-1, Naka, Kunitachi,
  Tokyo, 186-8601, Japan. Email:t.toriyabe@r.hit-u.ac.jp}}
\date{}

\begin{document}
\maketitle
\begin{abstract}
This study analyzes the gender gap in desired wages using large administrative data of public job referrals, which allows us to look at the desired salaries of individuals from a wider wage distribution. We conduct a decomposition analysis using available information on age, desired work region, and desired occupation. We find that of the three factors, desired occupation is the most important in generating differences in desired wages; however, the residuals are the largest outside of the three factors. To further probe the unexplained residuals, we also conduct heterogeneity and sensitivity analyses using the available data. ~
\end{abstract}

\newpage

\doublespacing

\section{Introduction}

The issue of the wage gap between men and women persists despite ongoing
efforts. Since the work of Albrecht, Björklund, and Vroman (2003),
advancements in quantitative analysis methods have revealed that the
wage gap does not exist uniformly; rather, the differences are more
pronounced among certain groups. In addition to the well-known glass
ceiling wage gap prevalent in management and higher positions, there is
also a concerning phenomenon known as the sticky floor effect, which
particularly affects lower-wage brackets; these patterns have been
observed in various countries (Arulampalam, Booth, and Bryan 2007; De la
Rica, Dolado, and Llorens 2008; Nicodemo 2009; Christofides, Polycarpou,
and Vrachimis 2013). Japan, which ranks among the developed countries
with the largest gender gap\footnote{According to the World Economic Forum (2025), Japan’s Gender Gap Index (GGI) in the 2025 edition stands at 66.6\%, ranking 118th out of 148 countries. This represents not only a low level in global terms but also falls well below the high-income country group average of 74.3\%.}, is no exception. The gender wage gap
persists even after accounting for factors such as workers' human
capital, and it tends to be more significant at both the top and bottom
ends of the wage distribution (Hara 2018).

Much of the research to date has analyzed data from workers who are
already employed. Analyses of the employed workers, who make up the bulk
of the population, suggest many social and policy implications. Yet, those
entering the workforce, especially those coming out of unemployment, may
have fewer years of experience in the firm and may be more likely to be
on the lower end of the wage scale. Indeed, the importance of focusing
on the extensive margin segment of the labor market to understand the
full extent of the gender gap has been pointed out (Olivetti and
Petrongolo 2008)\footnote{They note that women may tend to be left out
  of the sample for low-skill groups when it comes to understanding gender
  disparities. Importantly, however, although we focus on these women,
  our study does not address the issue Olivetti and Petrongolo (2008)
  raised because we do not endogenize their job search activities.}. If
people transition from unemployment, which comprises the weakest
position in the job market, but a wage gap persists between men and
women with similar characteristics, then women remain exposed to further
vulnerable economic circumstances.

In recent years, gender differences in job seekers’ wage preferences have attracted increasing attention as a potential factor contributing to the gender wage gap, with a growing body of research documenting differences in reservation wages and wage expectations between men and women (e.g., Brown, Roberts, and Taylor, 2011; Krueger and Mueller, 2016; Cortés et al., 2023).\footnote{See also Caliendo, Lee, and Mahlstedt (2017); Kiessling et al. (2019); Brown, Popli, and Sasso (2022); Basbug and Fernandez (2025); Bonaccolto-Töpfer and Satlukal (2024).} Within this line of work, a notable contribution is Roussille (2024), who used actual information from an online job platform to examine ask salaries and showed how such differences translate into contract wages. While her study provides valuable new insights, its scope is limited, as it focuses exclusively on high-skilled individuals in the cloud IT sector.

The purpose of this study is to estimate the gender gap in wage preference among those in broader groups of the wage distribution using
data from the large administrative dataset of public job referrals,
namely, Japan's Employment Security Office (PESO). Our data selection will
allow us to analyze high-quality data that cover a wider range of jobs,
including both high- and low-skill job vacancies, whereas prior studies
have concentrated on high-skill jobs. Specifically, our analysis focuses on desired wages, a measure that blends elements of existing concepts of wage preferences (Details are provided in \ref{sec-data}). The empirical analysis takes full
advantage of the uniqueness of the data; while the number of
individual-level variables is limited, the sample size is large. 

By
targeting broader groups in the wage distribution, the analysis helps
clarify the role of the gender gap in desired wages in a way that complements
previous studies. Moreover, compared with cloud IT jobs, another
potential strength of our data is the addition of a regional dimension.
By definition, cloud jobs are not affected by local labor supply and
demand, whereas service sectors, for example, are more local in nature;
therefore, it might be interesting to examine the role of discrimination
and monopsony in the gap in desired wages. It should also be noted that in the Japanese context, directly negotiating wages with companies is a relatively uncommon occurrence compared to other countries. Moreover, the ``ask salaries'' in this dataset are provided as references for job opportunities introduced by the employment security office. As a result, these desired wages do not reflect strategic incentives to negotiate with employers and can thus be considered relatively close to the individuals' own preferences.

The analysis results reveal the presence of a gender gap in desired wages,
highlighting a substantial value of 17.6 percentage points compared with
the value of 6.6 percentage points found among the skilled working
population in the U.S, as documented by Roussille (2024).\footnote{Note that Roussille (2024) analyzes \textit{asking wages (ask salaries)}, which are the wages that job seekers propose when submitting applications on the platform. In contrast, our study focuses on \textit{desired wages}, i.e., the reservation-like wages that job seekers report when registering at the public employment service. We explain this distinction in more detail later.}
 While this may be a country-by-country
difference, it is also consistent with the results of previous heterogeneity
analyses, which highlights a larger difference for the unemployed in the U.S.
(as we discuss in Section~\ref{sec-background}, the users of public job referrals in Japan include a large number of unemployed workers). We further conduct a
decomposition analysis to see how gender differences in the available
variables of age, desired region, and desired occupation affect the ask
gap. We find that among the three factors, differences in desired
occupation contribute significantly. However, the findings also reveal
that the unexplained factor is more prominent than the other three
factors, with a value of 14.8 percent points. To comprehend the
significance of the substantial residuals, we conduct
heterogeneity analyses and discover a significant level of
heterogeneity, prompting a discussion about the ask gap. Sensitivity analysis further shows that while the residual appears large, its magnitude is about one-third of the contribution of desired occupation, indicating that it is important but not the dominant driver of the gap.

Our results contribute to the literature that seeks to understand the
gender gap through wage preferences in various ways. First, our results show
that the ask gap exists despite the presence of a social context in
which wage bargaining is unlikely to occur (as explained in
Section~\ref{sec-background}). Second, we conduct a decomposition and
heterogeneity analysis of the factors that cause the gap. The
decomposition analysis reveals that the most explanatory factor that can
be observed is the difference in the distribution of occupations. The
heterogeneity analysis also reveals that when controlling for other
observable factors, the gap widens in middle age, and that women's
desired wages are lower even in female-specific jobs. Given these
results, for a wide range of occupations not restricted to high-skilled
jobs, we find indirect evidence that gives a new interpretation of the
gender gap by the compensation wage hypothesis or reference wage hypothesis
rather than the differences in wage bargaining attitudes and skills.

The rest of the paper is organized as follows.
Section~\ref{sec-background} presents the institutional setting in
Japan. Section~\ref{sec-data} explains the data we used, while
Section~\ref{sec-descriptive_statistics} provides descriptive
statistics. Section~\ref{sec-estimation_strategy} presents the
estimation strategy. Section~\ref{sec-results} presents the main
results, while Section~\ref{sec-heterogeneity} presents the
heterogeneity analysis. Section~\ref{sec-sensitivity} conducts the sensitivity analysis. In Section~\ref{sec-discussions}, we discuss
the results, and Section~\ref{sec-conclusion} concludes the paper.

\section{Institutional setting}\label{sec-background}

\subsection{Gender wage gap and Japanese employment
system}\label{gender-wage-gap-and-japanese-employment-system}

The wage gap between men and women in Japan has been gradually
decreasing over the long term.  In 2021, the average wage for female regular employees was 75.2\% of that of their male counterparts. However, when compared with the OECD countries' average of 88.4\%, the
gender wage gap in Japan remains relatively large in terms of
international standards.

Existing studies on the gender wage gap in Japan have highlighted its
connection to the country's distinctive employment system\footnote{For
  an excellent survey on the relationship between the gender wage gap
  and Japanese employment system, see Hara (2018).}, including features
such as lifetime employment, the seniority wage system, and the initial
promotion system (A. Kawaguchi 2015). Moreover, the nonperformance-based
wage system has also been identified as a contributing factor (Chiang
and Ohtake 2014). However, it is worth noting that these studies
primarily concentrate on employed workers and utilize realized wages as
the basis for their wage information. Thus, they do not provide insights
into ask salaries.

An essential aspect of the Japanese labor market when discussing salary
expectations is the method of wage determination upon entry into the
workforce. Developed countries such as the U.S. often rely on wage
negotiations between employers and job seekers (Hall and Krueger 2012);
however, this practice is not widespread in Japan. According to a survey
conducted by the Recruitment Work Institute in 2020, which focused on
job changers, a significant majority (58\%) reported accepting the wage
offered by their employer, which is in notable contrast to practices in
other developed nations\footnote{This report summarizes the results of a
  survey of college graduates in their 30s and 40s working in the
  private sector in Japan, the United States, France, Denmark, and
  China. In the other countries, less than 30\% of the respondents
  reported accepting the wage offered by their employers, and many
  reported telling their employers what they wanted to be paid.
  (Recruit Works Institute, 2020)}.
Consequently, from an economic standpoint, wage posting, rather than
bargaining, is the prevalent mode of wage determination in Japan. This
implies that job seekers typically have the choice of either accepting
or rejecting the wage proposed by the employer.

\subsection{Public employment security
office}\label{public-employment-security-office}

The data utilized in this study, including information on ``desired wages,'' is sourced from job application forms submitted to public
employment security offices. These offices, which number 544 across
Japan, offer three main services, namely, job placement, employment
insurance, and employment measures such as corporate guidance and
support. In 2022, these offices handled a total of 4,586,000 cases, with
10,052,800 new job openings and 1,226,000 job placements\footnote{Ministry of Health, Labour and Welfare (2020a).}.

According to a 2020 survey conducted by the Ministry of Health, Labor,
and Welfare, job changers employ various methods in their jobs search
activities\footnote{Ministry of Health, Labour and Welfare (2020b)}.
The results reveal that 38.4\% utilize ``job websites, job information
magazines, newspapers, flyers, etc.,'' while 34.3\% rely on ``public
institutions such as PESOs,''
and 26.8\% seek opportunities through ``nepotism.''

While private employment agencies generate their profits through
commissions and introduction fees from employers, PESOs do not charge such fees. PESO serves as a safety net for employment, with the government offering
free support, primarily targeting individuals who are facing challenges
in securing employment through private agencies. It also extends
assistance to small, medium, and microsized enterprises that are
experiencing a shortage of labor.

Therefore, importantly, users of PESO are not necessarily highly
skilled or advantaged in the labor market. Additionally, the sample
population includes unemployed individuals engaged in off-the-job
searches, as they are required to follow specific procedures, such as
applying to a PESO to qualify for unemployment insurance benefits.

Finally, we utilize the Labor Force Survey and the Survey on Employment Trends to analyze the characteristics of PESO users, who constitute the primary audience for the data presented. The Labor Force Survey, which focuses on individuals participating in the labor market, provides insights into ``job seekers,'' shedding light on how unemployed individuals search for jobs. Conversely, one of the main objectives of the Survey on Employment Trends is to explore how job seekers secure employment within Japan's major industries. Herein, we use the 2019 data in our main analysis.

The 2019 Labor Force Survey shows that 30.2\% of job seekers rely on public employment security offices. According to the 2019 Survey on Employment Trends, 9.9\% of underemployed individuals (excluding new graduates) and 16.7\% of job changers find employment through these offices. Among these job seekers and job entrants, an analysis by age reveals a low usage rate among younger individuals, while an analysis by company size indicates a low usage rate among those entering large companies.

Examining the characteristics of job seekers and the companies they join, it can be inferred that public employment security offices play a vital role in facilitating and supporting the matching of individuals disadvantaged in the labor market. This is evidenced by their primary users being older individuals, who face relatively greater challenges in changing jobs in Japan; it is also supported by the fact that the companies involved are generally small in size.

\section{Data}\label{sec-data}
\subsection{Analytical sample based on public employment service data}

We use data from job application forms submitted by individuals seeking full-time employment through Japan's public employment services between January and December 2019. These forms represent the first step in the public job referral process and serve as the basis for subsequent consultations with employment counselors. While the form asks for detailed information about each job seeker, our dataset includes only a subset of the collected items: gender, age, preferred work location, and preferred occupation.

We focus on 2019 data for two reasons. First, it is the most recent year unaffected by the COVID-19 pandemic. Second, as shown in Appendix B, there is evidence of monthly seasonality in job search activity, which differs by gender. Therefore, using a full calendar year is preferable to relying on data from specific months. 

Our dataset contains some missing values in the preferred occupation variable. Such cases were dropped from the analysis.
The final sample includes 2,905,952 individuals seeking full-time employment.

\subsection{Variables}
This subsection describes the variables used in the analysis, with a focus on the desired wage, which is interpreted in light of the institutional context and compared with existing economic concepts of wage preferences.
\subsubsection{Description of variables}
\textbf{Desired wage}

The desired wage is the main outcome of interest in our analysis. The form asks full-time job seekers to report the minimum monthly income they would accept, stated as "X yen or more." We use this reported amount as the lower bound of their desired wage (We discuss the interpretation of desired wage in this institutional setting in more detail later.) While the form also collects hourly wage data from part-time job seekers, our analysis focuses exclusively on full-time employment.

\textbf{Desired occupation}

Job seekers are asked to indicate their preferred occupation. Responses vary in specificity: some provide detailed five-digit codes, while others give broader occupational categories or leave the item blank.
The specific embedding method used for modeling is described in Section~\ref{sec-embedding-method}.

\textbf{Desired work location}

Applicants specify their preferred work location at the prefecture level. There are no missing values in this variable.

\textbf{Demographic characteristics}

The dataset includes basic demographic variables, such as gender and year of birth. Age is calculated using the year of birth and the time of registration.

\subsubsection{Institutional context and interpretation of desired wage}

The desired wage is a required field that all job seekers must complete when registering with the PESO. Accordingly, every registrant is obliged to provide this information, which is then used in the job-matching process conducted by PESO. While applicants are free to enter any amount at the time of registration, the reported figure is reviewed during the initial interview with PESO staff. If the amount is deemed excessively high or low, staff may advise the applicant to revise it. However, unless the value is clearly unrealistic, staff intervention is generally limited, and the submitted amount is often accepted as is.

What is important here is the specific characteristics of the job seekers who use PESO. Compared to those who use private-sector job platforms to pursue career advancement, PESO users tend to take a more cautious and realistic approach to job searching. In fact, it is not uncommon for applicants to underestimate their market value and report desired wages that are too low to maintain a basic standard of living. As a result, when such underreporting occurs, PESO staff may help applicants reassess their needs based on household composition and essential living expenses, often encouraging upward revisions of their applications. Through this process, the final desired wage tends to reflect not only personal preferences, but also a more realistic estimate of the income needed to maintain a basic standard of living.

This interpretation is consistent with the views expressed by PESO staff during interviews. According to them, the desired wage is not so much a reflection of the applicant’s perceived market value or wage expectations, but rather should be understood primarily as a practical minimum acceptable amount—one that takes into account the applicant’s living conditions and institutional guidance. For further details based on interviews with PESO staff, see Fukai et al. (2025).

\subsubsection{Comparison of our wage variable with existing concepts of wage preferences}

Building on the institutional context outlined above, we now situate the desired wage variable used in our analysis within the broader literature on wage preferences. Specifically, we compare it with three well-known concepts—reservation wage, expected wage, and ask salary—across four key dimensions: definition and determinants, observability and use in matching, method of data collection, and susceptibility to strategic behavior.

\textbf{Definition and determinants}

The reservation wage represents the minimum acceptable wage a job seeker is willing to receive and is shaped by personal constraints and preferences, such as financial needs, family responsibilities, and alternative options. The expected wage reflects an individual's subjective belief about their likely earnings in the near future, incorporating both optimism and market expectations. The ask salary, by contrast, is often strategic in nature—used as a negotiation anchor or a signal to employers. In our case, the desired wage reflects both individual preferences and practical considerations, including minimum living costs. It is not purely subjective or aspirational, but grounded in a concrete institutional process that encourages realism.

\textbf{Observability and use in matching}

Reservation and expected wages are typically private and not disclosed to employers or intermediaries. Ask salaries are explicitly stated and visible to firms, often forming a basis for negotiation. Our variable occupies an intermediate space: it is required for registration with PESO and used for job-matching purposes, but it is not shared with employers. As such, it informs the job search process without serving as a direct signaling device.

\textbf{Method of data collection}

Reservation and expected wages are commonly obtained through surveys or experiments \footnote{Reservation wages are most commonly measured via surveys or experimental methods to explore their gender difference (Brown et al. 2011; Caliendo et al. 2017; Brown et al. 2022; Cortés et al. 2023; Basbug and Fernandez 2024; Bonaccolto-Töpfer and Satlukal, 2024). In the case of France, however, Le Barbanchon et al. (2021) have also collected them directly through reports submitted to the public employment service. Recent studies on expected wages are, on the other hand, often based on surveys or experiments conducted among university students (Blau and Ferber, 1991; Brunello et al., 2004; Reuben et al., 2017; Kiessling et al., 2019).}, while ask salaries are observed in online job platforms or negotiation settings. For instance, Roussille (2024) analyzes ask salaries using data from a large online hiring platform for IT jobs, where candidates publicly specify their desired wage as part of the application process. In contrast, our variable is collected through a mandatory administrative process. Job seekers input their desired wage during registration, and the value may be revised through interviews with PESO staff. This procedure enables consistent and standardized data while preserving flexibility in applicant input.

\textbf{Gender differences in wage preference measures}

Prior research has consistently documented gender gaps across various wage preference measures, including reservation wages, expected wages, and asking salaries. Women tend to report lower values than men, and these tendencies have been suggested to contribute to the actual gender wage gap. These differences have been attributed to a range of factors, including differences in occupational choices (Basbug and Fernandez, 2025), family responsibilities (Blau et al. 1991; Brown et al. 2011; Le Barbanchon et al., 2021; Bonaccolto-T{\"o}pfer and Satlukal, 2024), and commuting constraints (Le Barbanchon et al., 2021; Brown et al., 2022), as well as behavioral traits such as lower self-assessed market value, risk aversion, and a greater tendency to avoid conflict or assertiveness in negotiation contexts (Caliendo et al., 2017; Reuben et al., 2017; Kiessling et al., 2019; Cortés et al., 2023; Roussille, 2024). These behavioral patterns may cause women to report lower wage expectations or reservation wages, not necessarily because of lower productivity expectations, but due to social norms and internalized beliefs about appropriate wage demands. As a result, the interpretation of gender gaps in these measures often requires careful consideration of strategic behavior and socialized preferences.

\textbf{Potential for strategic bias in reported wage preferences}

The extent to which wage preference measures are susceptible to strategic reporting largely depends on how those measures are used in practice. For instance, as discussed above, ask salaries in Roussille (2024) are typically visible to employers and serve as a starting point for wage negotiations. As such, job seekers have an incentive to report a higher value than their actual minimum. This is because stating one’s true desired wage may lead to downward adjustments during negotiation, potentially resulting in an unfavorable outcome. In such contexts, reporting a slightly inflated value is often considered a rational bargaining strategy.

By contrast, other wage preference measures, including the desired wage used in this study, as well as the more conventional reservation and expected wages, are not disclosed to employers and are not directly involved in wage bargaining. These figures may affect matching decisions, but they are not used as reference points in negotiations. Accordingly, the incentive for strategic inflation at the time of reporting is less explicit.

That said, these measures are not entirely immune to the influence of strategic behavior. In particular, internal measures such as reservation and expected wages may reflect cognitive and behavioral traits related to negotiation, such as risk aversion, confidence, or attitudes toward competition and conflict. Prior studies have shown that these traits can systematically influence reported values and that such patterns often differ by gender (Caliendo et al., 2017; Reuben et al., 2017; Kiessling et al., 2019; Roussille, 2024). These findings suggest that reported wage preferences may not always reflect productivity expectations or market forecasts, but may instead be shaped by individual-level behavioral dispositions.

In this context, the desired wage variable used in this study is considered relatively less susceptible to strategic behavior due to the institutional setting in which it is collected. As discussed in Section~\ref{sec-background}, wage negotiation is uncommon in the Japanese labor market, particularly among users of the Public Employment Service Offices (PESO). Job seekers are asked to report their desired wage at the time of registration, often guided by practical needs and institutional counseling. The reported figure is not shared with employers, which reduces the incentive to manipulate the value for signaling purposes strategically.

To be clear, the desired wage in our data is not equivalent to existing economic concepts such as the reservation wage, expected wage, or ask salary. It is an administratively collected figure, shaped by institutional procedures and practical constraints. Nonetheless, because it is not used in bargaining and is less exposed to strategic considerations, it provides a useful window into individual preferences and constraints. In this paper, we use this variable to examine gender differences in wage preferences under an institutional setting that arguably minimizes the confounding effects of strategic behavior.

\subsection{Embedding method}\label{sec-embedding-method}

In the analyses, the embedding method is used, a dummy for missing
values is created, and the median of the embedded values is substituted.
Generating many dummies is generally inefficient in terms of estimation
and increases the computation time (indeed, it is quite a serious
problem since the sample exceeds 500,000 cases). Therefore, the
following two methods are used to ``embed'' them in a two-dimensional
vector space. First, we estimate and use predicted wages (the lower
bound of the offered wage) from job postings by both prefecture and by
occupation. For occupations with many categories, LASSO estimation is
used. Then, we compute and use the means of each category for the other
variables in the data (here, only age). It has been previously argued that this
method can be justified as sufficient representation (Johannemann et al.
2019).

\section{Descriptive statistics}\label{sec-descriptive_statistics}

In this section, we present the average gap in desired wages and the ratio of female job seekers for each descriptive statistic group using the analyzed data.
To divide these groups, we first estimate each region and occupation's average wage (lower bound) using the job offer data. Note that, for ease of interpretation, we refer to the offered wage for males to construct the groups.

These offered wages are obtained from job posting records, which constitute another administrative dataset maintained by PESO as part of its core operations. Employers report monthly salary ranges (both lower and upper bounds) when registering job openings, and these values are entered in yen units. For our analysis, we focus on the lower bound of the offered wage range.

Then, groupings are made via quintiles for the regions (5 groups of regions) and occupations (5 groups of occupations + nonrespondents), and quintiles are also used for age (5 groups of ages). The result is a total of 150 ($=5 \times 6 \times 5$) groups with average desired wages and percentages of women according to gender. The following is a description of each of the groups classified by quintile.

\subsection{Classification of groups}\label{classification-of-groups}

First, age in the job seeker form is divided into five quantiles. For the regional groups, the
fifth quantile of 80--100\% includes two prefectures in the Tokyo
metropolitan area, while the fourth quintile of 60--80\% includes Osaka,
Aichi, and other metropolitan centers, indicating that the larger the
the population size of a prefecture is, the higher the offered wage is.
Finally, the quintiles of average offered wages by desired occupations
obtained from the job postings and the groups that did not answer the
desired occupations are classified.

\subsection{Wages - background
attributes/gender}\label{wages---background-attributesgender}

\begin{figure}[H]

\centering{

\includegraphics{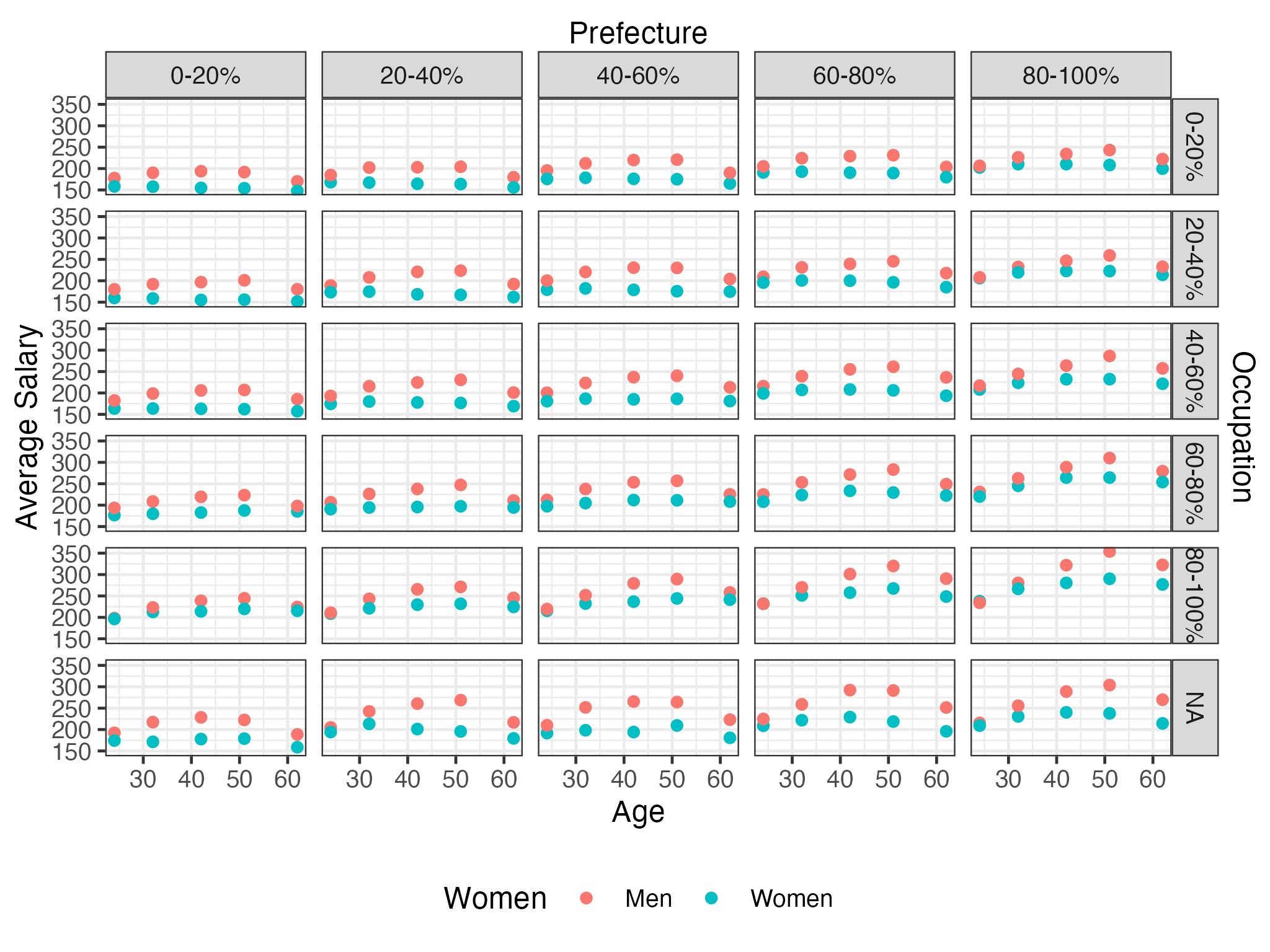}

}

\caption{\label{fig-Figure-askwage-by-quantile}The ask salaries of men
and women}

\end{figure}

Figure~\ref{fig-Figure-askwage-by-quantile} shows the desired wage gap by
quartile according to the background attributes presented above in a
single figure (for the overall distribution of desired wages by gender,
see Appendix A). The way to look at this figure is as follows. The
labels on the upper side line up the quintiles of the regions spanning
from left to right. The labels on the right side line up with the
occupational quintiles running from top to bottom. The horizontal axis
at the bottom of the figure shows the age quintiles within each quintile
of the region. Finally, the vertical axis in each cell on the left side
shows the average desired wage in units of 1,000 yen for each subgroup.
The red dots denote males, and the blue dots denote females. From this
figure, we can see the following.

Let us look at each characteristic separately. First, focusing on the
regional part, the desired wages are higher according to the offered
wage. Second, by occupation, the difference in the level of desired wages
between men and women is small for occupations with low-wage offers;
however, the difference tends to be large for occupations with high
offered wages. A large difference is also observed in the group with no
desired occupation. Third, in terms of age, a gender difference is
observed in the middle-aged group.

In summary, these characteristics show no gender differences across
regions or occupations with low-wage offers (upper left). This finding
implies that the minimum wage could influence individuals' desired wages,
leading to a narrowing of 
the gap; this is similar to the phenomenon
whereby the minimum wage can narrow the gender (actual) wage gap through
the change in the wages of those in the lower wage distribution (Blau
and Kahn 1997). On the other hand, the gender difference opens up when
the offered wages tend to be high (toward the lower right), and a large
difference is observed, especially in the middle-aged group.

\subsection{The ratio of women to men - background
attributes}\label{the-ratio-of-women-to-men---background-attributes}

\begin{figure}[H]

\centering{

\includegraphics{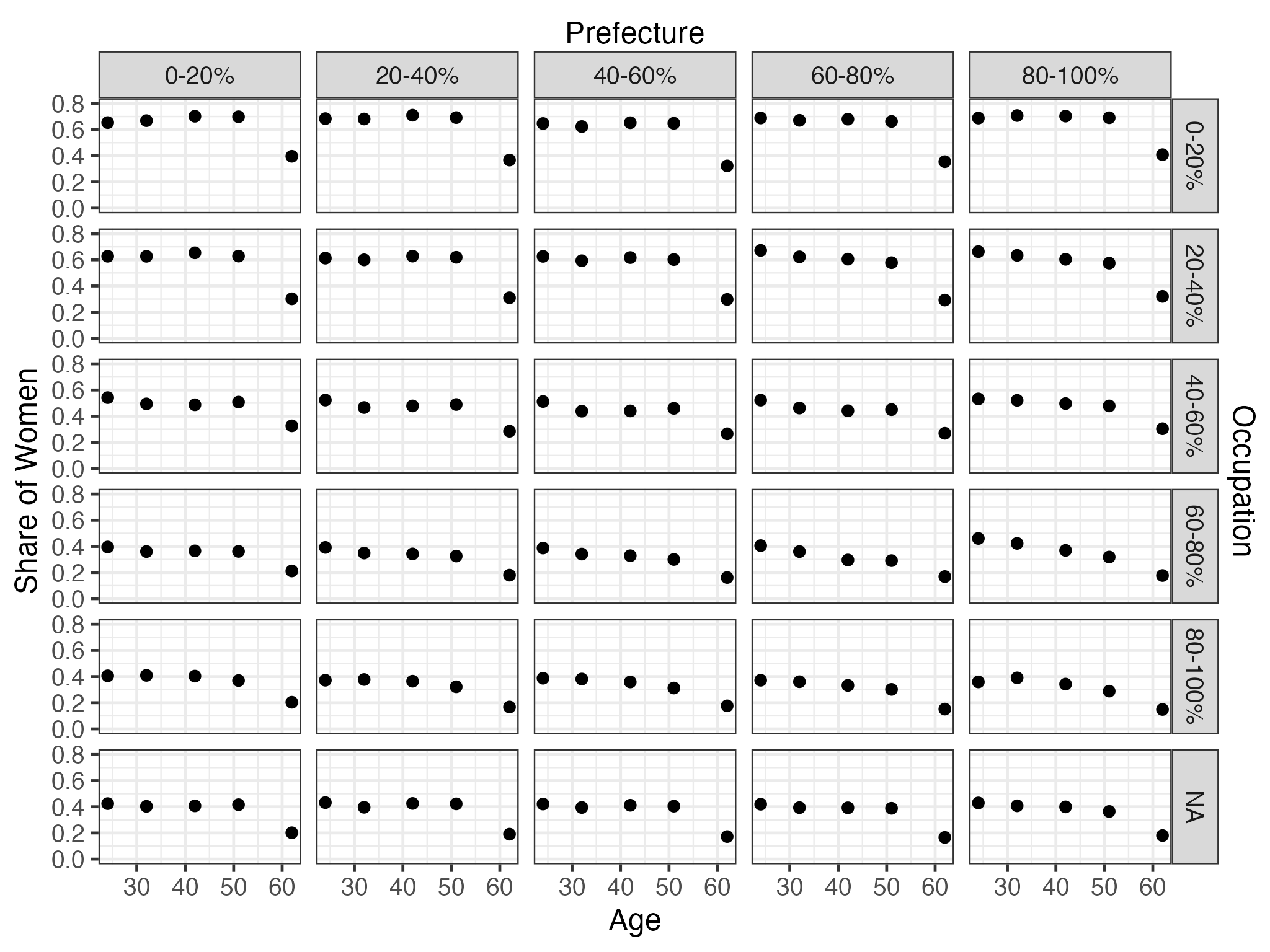}

}

\caption{\label{fig-Figure-female-ratio}The ratio of women to men}

\end{figure}

This section shows the ratio of women within each background attribute
to facilitate the interpretation of the generalized KOB-Duncan method, which is used in Section \ref{sec-estimation_strategy}.
Each plotted point in Figure~\ref{fig-Figure-female-ratio} shows the
ratio of women to men in the corresponding group; at a value of 0.5, the
ratio of men to women is the same. Figure~\ref{fig-Figure-female-ratio}
shows the following.

First, there are no large differences by desired region. Second, the
ratio of women declines in desired occupations with higher offered
wages. Third, there is a sharp decrease in the ratio of women in the
fifth quantile of age.

\subsection{Comparison with the actual wage in the labor
market}\label{comparison-with-the-actual-wage-in-the-labor-market}

In this subsection, we aim to elucidate the discrepancies between the
desired wage gathered through PESO and the actual labor market wages.
The Ministry of Health, Labor and Welfare (MHLW) annually conducts the
Basic Survey on Wage Structure, which is a statistical survey that aims
to elucidate the real wages of workers across major industries. They
also publish aggregate data on monthly salaries categorized by
employment status, gender, age group (in 5-year intervals), and other
demographics; thus, we use this data to draw comparisons with the desired wages of our sample.

\begin{figure}[H]

\centering{

\includegraphics{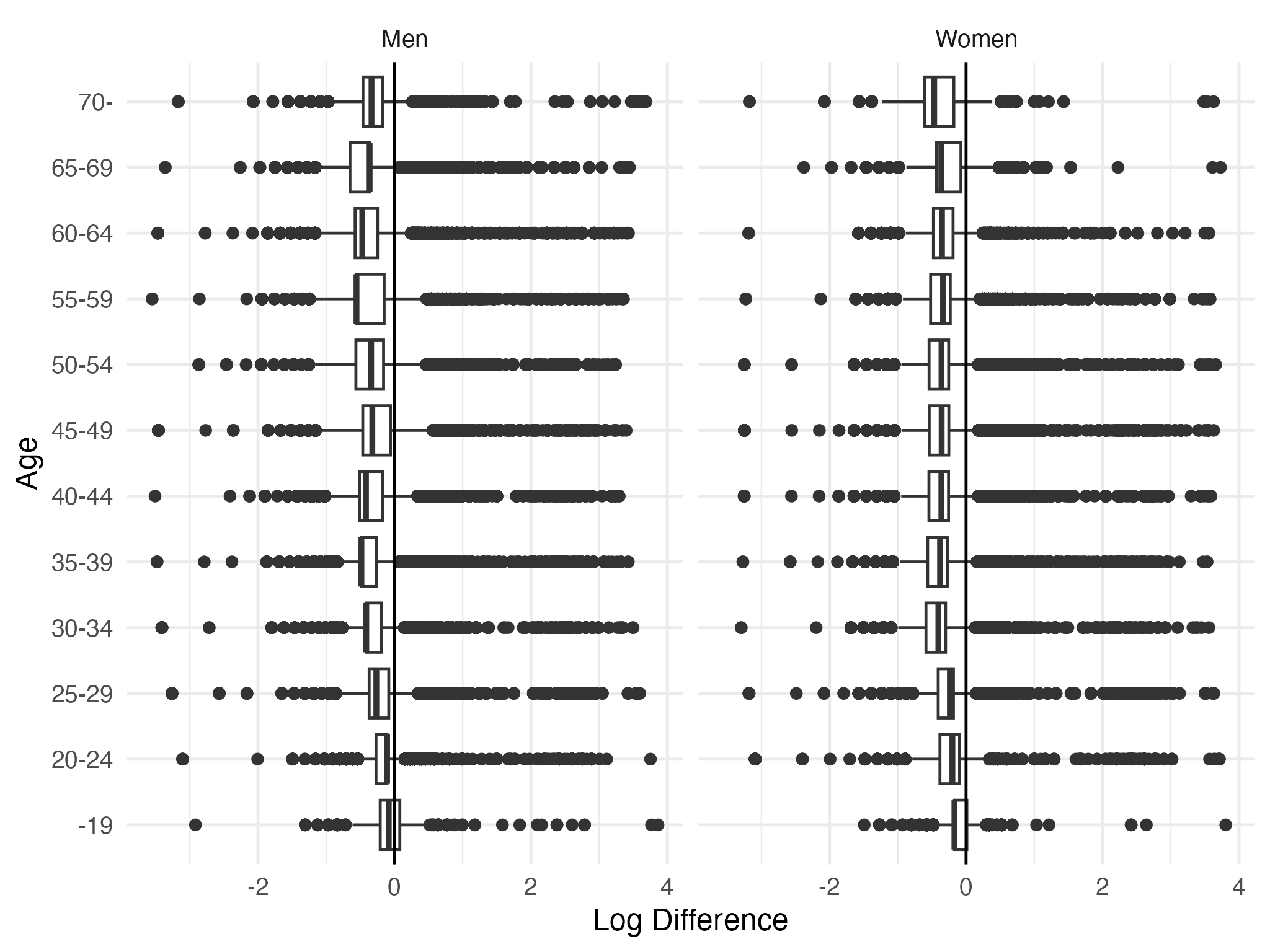}

}

\caption{\label{fig-Figure-comparewith-actualwage}The comparison with
the actual wages (all sample)}

\end{figure}

To understand the relationship between wages in the actual labor market
and the desired wages in our specific sample,
Figure~\ref{fig-Figure-comparewith-actualwage} illustrates a
box-and-whisker plot. This visual representation highlights the
difference between the logarithm of the average wage for full-time
workers in 2019, categorized by gender and age group (as per our
sample), and the logarithm of the desired wage for individual job applicants
within the same gender and age group. Specifically, it illustrates the
logarithmic value of each worker's desired monthly salary minus the
logarithmic value of the average actual monthly salary.

Figure~\ref{fig-Figure-comparewith-actualwage} indicates that the median
value consistently falls to the left of 0 across all groups. This
suggests that the desired wages reported at PESO tend to be lower
than the wages observed in the labor market for individuals with similar
demographic characteristics. This finding is consistent with the context of the current study in not only that the job seekers in our data mainly consist of those who are unemployed but also that their desired wages serve as their reservation wages.

\begin{figure}[H]

\centering{

\includegraphics{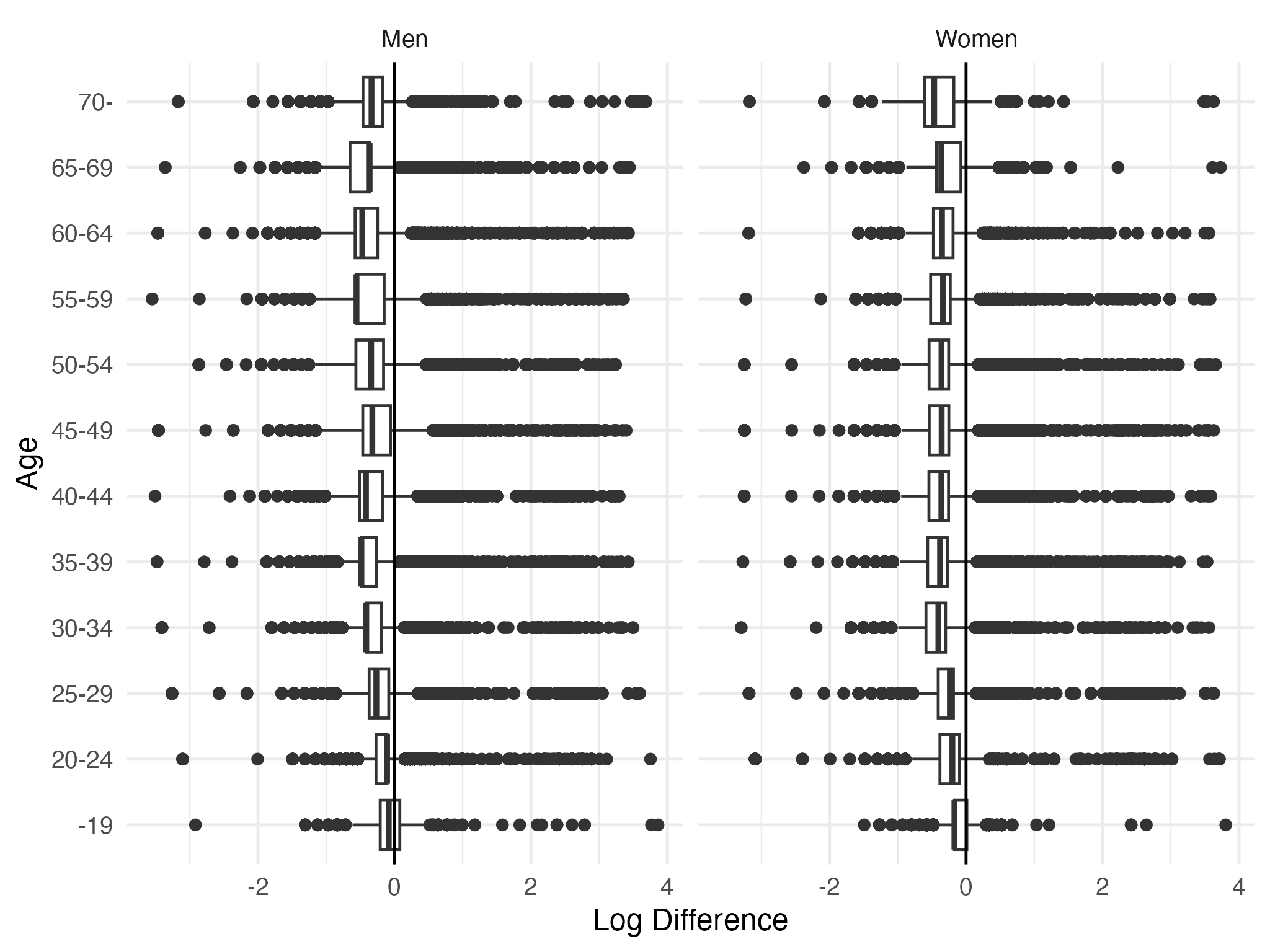}

}

\caption{\label{fig-Figure-comparewith-actualwage-0}The comparison with
the actual wages (sample of workers with zero years of tenure)}

\end{figure}

Nonetheless, it is important to consider that the actual average of
salaries incorporates individuals with longer tenure; this factor may
contribute to the difference in the desired wages of job seekers. To
address this issue, we further examine the average wage of workers with
zero years of tenure, which is a group that is more closely aligned with
the job seeker's average wage.
Figure~\ref{fig-Figure-comparewith-actualwage-0} presents a
box-and-whisker plot illustrating the disparity between the logarithm of
each job seeker's desired monthly salary and the logarithm of the average
monthly salaries for workers with zero years of tenure, categorized by
sex and age group (the latter minus the former as
Figure~\ref{fig-Figure-comparewith-actualwage}).

Figure~\ref{fig-Figure-comparewith-actualwage-0} suggests that even when
considering the actual wages of workers with zero years of tenure, their
desired wages continue to exhibit a tendency to be lower, which is
consistent with the observations highlighted in
Figure~\ref{fig-Figure-comparewith-actualwage}. However, it should also
be mentioned that while the gap in desired wages is lower than the actual wage, it is
not that far off in terms of average.

\section{Estimation strategy}\label{sec-estimation_strategy}

This study attempts to answer two main questions. The first question
focuses on the extent to which each observable attribute contributes to
the desired wage gap. The second question focuses on the extent to which
factors other than observable attributes contribute to the desired wage gap
between men and women. In the following subsections, we explain how to
answer each question.

\subsection{Decomposition}\label{decomposition}

To answer the first question, we first use the generalized KOB-Duncan
method. In decomposition analysis, the overlap assumption necessitates
that all the treatments of interest are observed in every subgroup.
However, as we discuss in Section~\ref{sec-descriptive_statistics},
there are subgroups where female applicants are not observed (for
example, older women are less likely to actively seek employment and are
virtually absent in certain occupations). Therefore, we only utilize the
subsamples that meet the positivity assumption criteria by excluding
subgroups with a male ratio exceeding 99.9\%. Decomposition analyses are
thus carried out by aligning the distribution of men with that of women.

The estimand in this analysis is the gap in desired wages between men and women
adjusted for the gender gap in covariates \(X\), which is defined as
follows:

\begin{equation}
\label{decomposition}
\phi(X)=\int_{X} \biggr[E[Y|D=1,X] - E[Y|D=0,X]\biggr]\times f(X|D=1)dX,
\end{equation}

where \(D=1\) means female and \(D=0\) means male, and \(f(X\textbar D=1)\) is
the distribution function of \(X\) in females. The above equation is
interpreted as the average gap in desired wages when the distribution of male
attributes is matched to that of female attributes.

The set of covariates \(X\) to be adjusted is extended sequentially from \(X_1\) to \(X_4\):

\begin{itemize}
\item
  \(X_1=(Month)\)
\item
  \(X_2=(Month, Age)\)
\item
  \(X_3=(Month, Age, Desired\ Region\ to\ Work)\)
\item
  \(X_4=(Month, Age, Desired\ Region\ to\ Work, Desired\ Occupation)\)
\end{itemize}

The above estimation allows us to identify within which covariates the
gender gap leads to the gender gap in desired wages. The estimation is carried out
using the moment conditions presented in Hahn (1998).

\subsection{Heterogeneity analyses}\label{heterogeneity-analyses}

Even after controlling for the month of employment, age, desired region, and
desired occupation, there is a possibility that a residual gender wage gap may remain. To deepen the understanding of the residual gender gap, we conduct two heterogeneity analyses. One is the best linear
predictor (BLP) on the residual gender gap in desired wages, and the other is the
subgroup analysis for pink-collar jobs on the basis of decomposition
analysis above (for the latter, see Appendix C).

To determine for which groups the gender gap is larger, we estimate a BLP of
\(\phi(X_4)\). Specifically, we estimate a linear approximate model
for \((Z=\) age, desired region to work, and desired occupation\()\). In
the analysis, the data are divided into quantiles for age, desired
region, and desired occupation (we add an unknown category for desired
occupation to create 6 quantiles). Dummy variables are then created for
each. When dividing into quantiles for the desired region and the
desired occupations, the average offered wage for the job is used.

Referencing Semenova and Chernozhukov (2021), the moment conditions are
as follows: \[0=\int Z (\phi(Z) - \beta Z)f(Z)dZ,\] where
\[\phi(Z)=\int E[Y|D=0,X]f(X|Z,D=1)dX\]
\[=\int E[Y|D=0,X]\frac{f(D=1|X)f(X|Z)}{f(D=1|Z)}dX.\] Using the Hines
et al. (2022) method, the estimates can be derived as follows:
\[0=\sum Z\Biggr(E[Y|D=0,X]\frac{f(D=1|X)}{f(D=1|Z)}\]
\[+(1-D)\frac{Y-E[Y|D=0,X]}{f(D=0|X)}\frac{f(D=1|X)}{f(D=1|Z)}\]
\[+E[Y|D=0,X]\frac{1}{f(D=1|Z)}(D-f(D=1|X))\]
\[-E[Y|D=0,X]\frac{f(D=1|X)}{f(D=1|Z)^2}[D-f(D=1|Z)]-\beta Z\Biggr).\]

The moment conditions require to estimate \(E[Y|d,X],f(D=1|X),f(D=1|Z)\) .

\subsection{Estimation of nuisance
functions}\label{estimation-of-nuisance-functions}

Both the decomposition and the BLP on the residual gender gap require
the estimation of several nuisance functions. Specifically,
\(E[Y|d,X],f(D=1|X),f(D=1|Z)\) should be estimated. In this paper, the
estimation is conducted by cross-estimation using LASSO, OLS, and random
forest stacking methods.

\section{Results}\label{sec-results}

\subsection{Average desired wage gap between men and
women}\label{average-ask-wage-gap-between-men-and-women}

\begin{figure}[H]

\centering{

\includegraphics{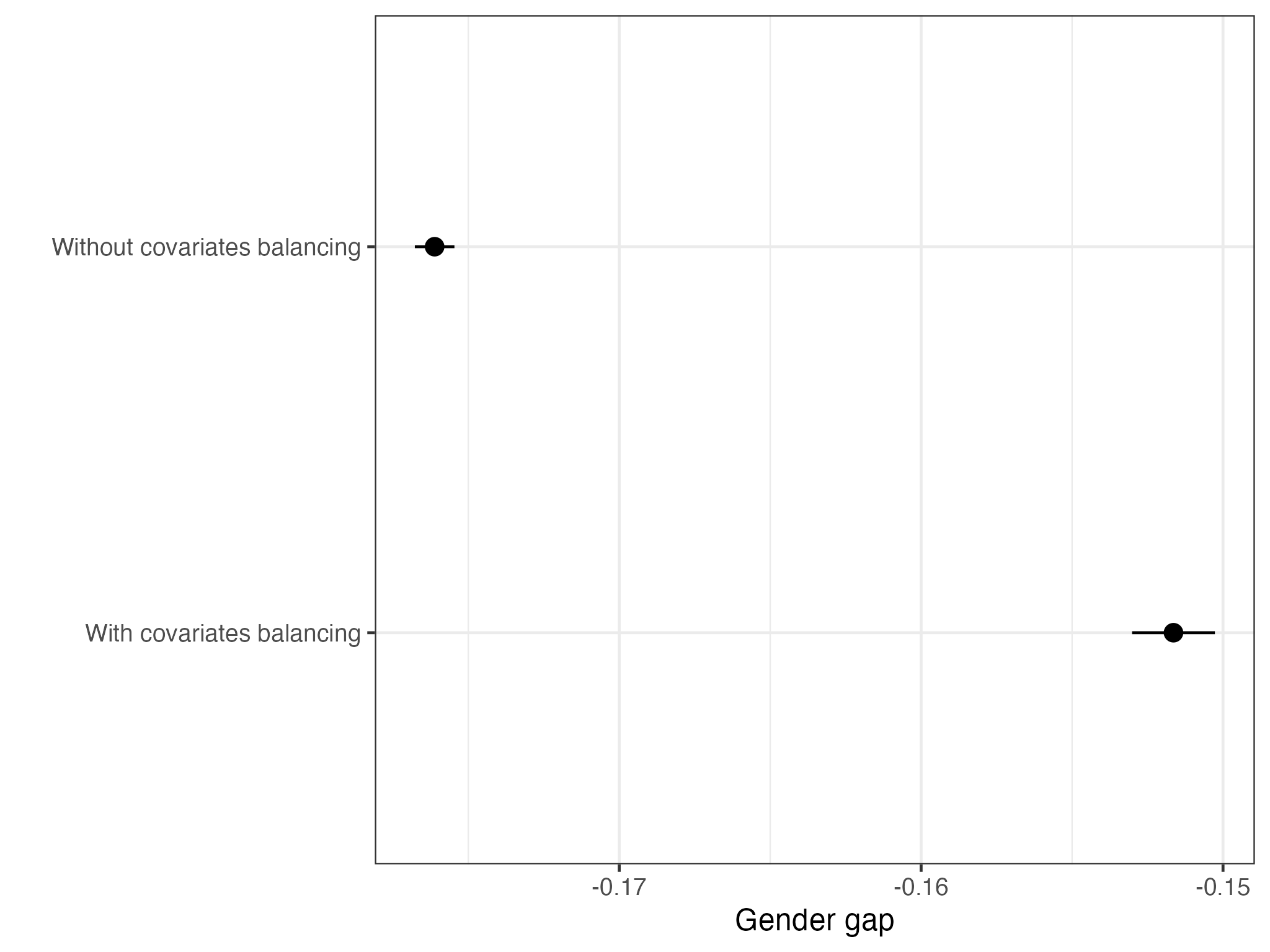}

}

\caption{\label{fig-Figure-raw-controlled-askgap}The raw desired wage gap
and the desired wage gap with all attributes controlled}

\end{figure}

Figure~\ref{fig-Figure-raw-controlled-askgap} shows the results of the
raw desired wage gap and the desired wage gap with the distribution of covariates balanced. The raw average wage desired for by women is 17.6
percentage points lower than that desired for by men, whereas the gap is reduced to 14.8 percentage points after controlling for age, region, and desired occupation.

Compared with Roussille (2024), who analyzed the ask wage gap for
skilled workers in the United States and found an overall raw ask wage
gap of 8 percentage points, the magnitude of 17.6 percentage points
found herein is quite large. Although this difference might be due to the presence of distinct labor market institutes in the U.S. and Japan, we suspect that it can be more attributed to the difference in the target populations since Roussille (2024) studied high-skilled job seekers while we study unemployed individuals who are presumably low-skilled job seekers. In fact, her
results show that the ask wage gap for unemployed workers is larger than
the average ask wage gap.

Let us now compare these results with the gender wage gap in Japan. It
has been noted that the average (actual) wage gap in Japan is 24.8
percent points (as discussed in Section~\ref{sec-background}). Compared
to this value, the difference in desired wages is 17.6 percentage points,
suggesting that the desired wage gap may have some effect on the actual wage
gap.

Finally, given the difference between the raw desired wage gap of 17.6
percent points and the conditional desired gap of 14.8 percentage points, one
might think the observed characteristics is not informative for the gender gap in desired wages. However, the observed characteristics of this difference may be
masked by counteracting factors. In the following subsection, we check
this by decomposition analysis.

\subsection{Decomposition analysis}\label{decomposition-analysis}

In this section, we perform a decomposition analysis of the gap in desired wages.
Specifically, we estimate Eq. (\ref{decomposition}) using the
generalized KOB-Duncan method. Recall that, as discussed in
Section~\ref{sec-estimation_strategy}, this analysis allows us to see
how the gap in desired wages changes when the distribution of male attributes is
aligned with that of female attributes.

\begin{figure}[H]

\centering{

\includegraphics{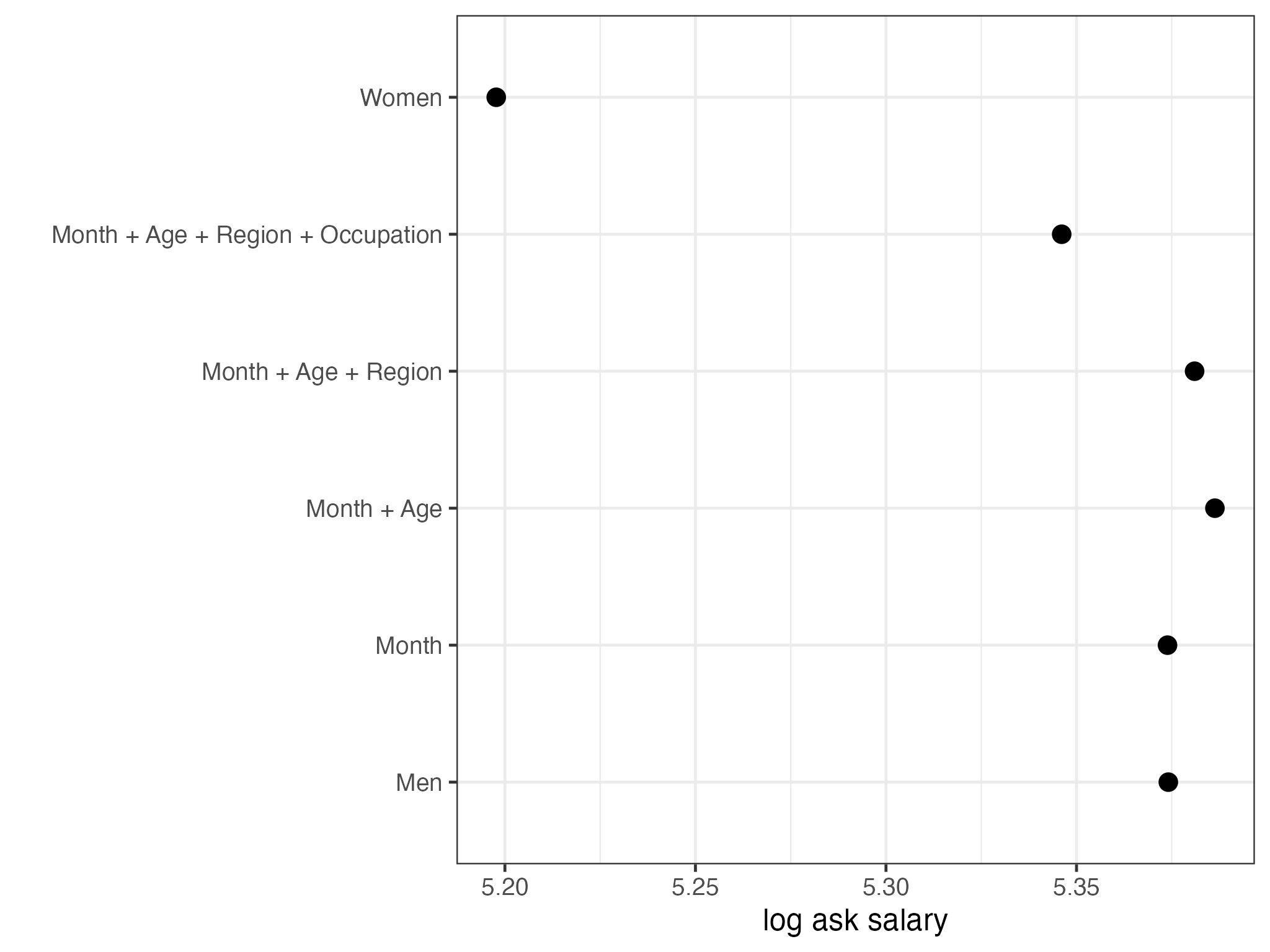}

}

\caption{\label{fig-Figure-decomposition}The results of the generalized
KOB-Duncan method}

\end{figure}

The results of the analysis are shown in Figure~\ref{fig-Figure-decomposition}. We can see that the desired wages for men decrease if the distributions of each factor of desired occupation and desired region are equal for women. The difference is largely due to the effect of the occupational category. The effect of the occupation is nonnegligible, accounting for  19.8  percent of the gender gap. Second, if the distribution of age is equal to that of women, then the male desired wages would increase by 7.1 percent. Thus, age is a factor that reduces the desired wage gap between men and women, while the desired region and desired occupation are factors that increase the gap. Finally, there is still a gap of approximately 15 percentage points such that the observed factors cannot explain.

The interpretation behind these results can be discussed as follows. As
seen in Figure~\ref{fig-Figure-decomposition}, the desired wage tends to be
lower in the older age group. However, as seen in
Figure~\ref{fig-Figure-female-ratio}, since the proportion of women is
higher in the lower age group, we expect the results to show higher desired wages for men when their age distribution is aligned with that of
women. Next, we consider why the desired occupation is the main factor
that widens the difference in desired wages. This may be because fewer women
apply for jobs that tend to have higher desired wages. Relatedly, it is
expected based on Figure~\ref{fig-Figure-female-ratio} that the ratio of
women is lower in jobs with higher offered wages.

In summary, as expected in the previous section, the differences in the
gender distributions of the two factors have opposite effects on the desired wage
gap such that they cancel each other out. Specifically, they are roughly
-20\% for desired occupation and +10\% for age compared with the overall
gender gap; thus, the distributional differences in desired occupation
and age are nonnegligible.

\section{Heterogeneity}\label{sec-heterogeneity}

It is clear that the residuals of factors other than age, desired
region, and desired occupation, i.e., residuals, are large with respect
to the desired wage gap between men and women. Unfortunately, due to data
limitations, it is not possible to directly identify these factors.
However, it is possible to estimate which combinations of age, desired
region and desired occupation have greater residuals among men and women
and to obtain indirect implications for analogizing the factors.

\begin{figure}[H]

\centering{

\includegraphics{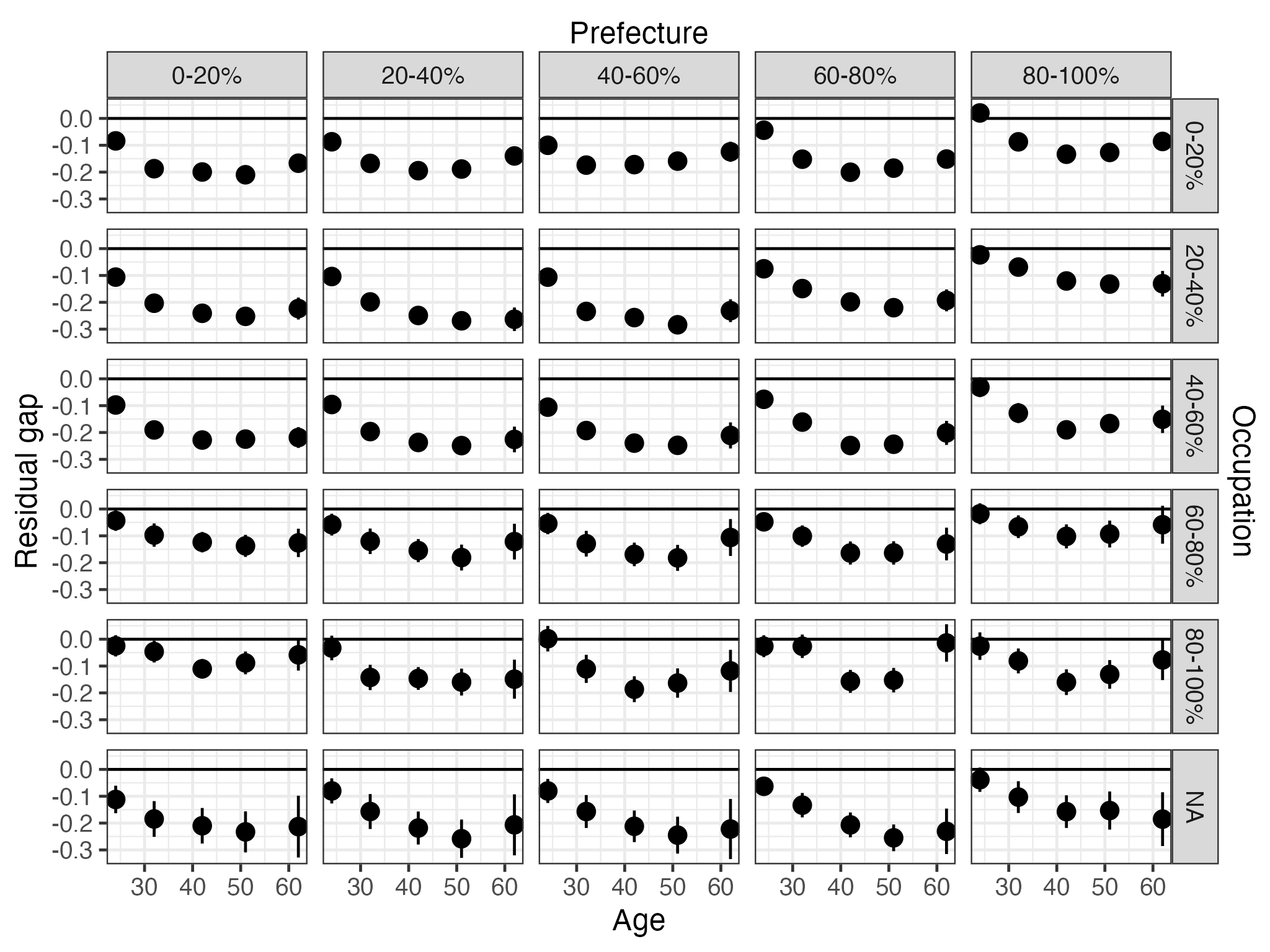}

}

\caption{\label{fig-Figure-cate150}Conditional average treatment effect}

\end{figure}

We perform a subgroup analysis to gain an intuitive understanding of the
heterogeneity within the residuals. Figure~\ref{fig-Figure-cate150}
shows the coefficients of the gender gap in desired wages for each
subgroup presented in Section~\ref{sec-data}. However, it is notable
that we control for age, desired region, and desired occupation in a
semiparametric way.

Several things can be learned from this figure. First, the residuals
differ markedly with respect to age. Specifically, as in the case of
labor supply, there is a U-shaped relationship with a drop-off in the
middle age. This may be due to the same reasons for the labor supply
(intrahousehold division of labor, career interruption for women).
Second, we see that the residuals are smaller among the younger
subgroup. Drawing on these heterogeneity results and the decomposition
analysis, we delve into a comprehensive discussion of the gender gap in
desired wages in the following section.

\section{Sensitivity analysis}\label{sec-sensitivity}
In our decomposition analysis using age, preferred occupation, and preferred region, we found that the unexplained component accounted for a very large share. This likely reflects the limited set of available variables, implying that unobserved factors may play a substantial role. Such factors may include gender differences in skills, work experience, or more detailed occupational preferences.

As a supplementary analysis, we conducted a sensitivity exercise following Chernozhukov et al. (2022). We used occupation as a benchmark variable, since it is expected to be strongly correlated with such unobserved characteristics. Specifically, we measured the relative importance of unobserved factors by comparing their hypothetical contribution to that of occupation in explaining the unexplained component.

This approach allows us to assess whether the unexplained component is of a magnitude comparable to occupational segregation. The results show that if unobserved factors related to skills, experience, or detailed preferences had an effect equivalent to about one-third of the contribution of occupation, the unexplained component could be fully accounted for. Since occupation itself is a broad and powerful predictor, this one-third threshold represents a substantial magnitude.

\section{Discussions}\label{sec-discussions}

This section synthesizes the empirical findings of our study to interpret the observed gender gap in desired wages and to explore the underlying mechanisms behind it. Our factor decomposition analysis reveals that among the three observable factors, the difference in the distribution of desired occupations is the most influential contributor to the gender gap. This is consistent with prior research on actual wages, which has long emphasized the role of occupational segregation (Blau and Kahn 2017). However, we also find that the unexplained residual—the component not accounted for by observed occupation, education, or experience—is even larger than the explained portion. Although this residual is not negligible, our sensitivity analysis indicates that its contribution is roughly one-third that of occupation (see Section~\ref{sec-sensitivity}), implying that additional mechanisms beyond observed characteristics may still be relevant.

To explore these mechanisms, we conducted heterogeneity analyses. The first analysis shows that the gender gap in desired wages is most pronounced among middle-aged job seekers. As seen in Figure~\ref{fig-Figure-askwage-by-quantile} and discussed in Section~\ref{sec-descriptive_statistics}, this pattern is driven primarily by rising desired wages among men, not declining desired wages among women. Crucially, these men are not employed but actively seeking jobs. In Japan’s job market, where internal promotions are still important, mid-career men may no longer expect upward mobility within firms and instead demand higher initial wages when switching jobs. Women, in contrast, may prioritize location stability or flexibility due to caregiving responsibilities.

These two patterns are consistent with the compensating wage differential hypothesis (Goldin 2014), which suggests that women are more likely to trade off higher earnings for non-wage attributes that facilitate work-life balance. Similar patterns have also been observed in studies of wage preferences. Blau and Ferber (1991) show that family responsibilities are linked to gender differences in expected earnings among business school students. Brown et al. (2011) find that gender gaps in reservation wages among parents are largely explained by the presence of children, while more recent studies attribute lower reservation wages among women to geographic constraints stemming from caregiving roles (Brown et al. 2022 for Italy; Le Barbanchon et al. 2021 for France). This interpretation is further supported by Basbug and Fernandez (2025), who show that gender differences in wage expectations among U.S. job seekers are largely driven by occupational preferences, which may themselves reflect broader social and institutional constraints. These findings suggest that women may be more willing to accept wage disparities based on the characteristics of their preferred jobs—a pattern that may also apply in Japan’s job posting-based search process.

The observed age-wage pattern also aligns with the reference wage hypothesis. In Japan, where wages are often determined by seniority rather than job-specific productivity, individuals may form expectations by comparing themselves to others with similar demographic characteristics, such as age and gender. Even unemployed job seekers may anchor their expectations to prevailing wage norms observed among peers. This mechanism helps explain why men’s desired wages increase with age while women’s remain flat. Fehr, Goette, and Zehnder (2009) provide experimental evidence that reservation wages are strongly shaped by reference points, while others have emphasized the psychological impact of relative income comparisons (Clark, Frijters, and Shields 2008; Godechot and Senik 2015; Noy and Sin 2021).
\footnote{Fehr, Goette, and Zehnder (2009) provide experimental evidence that reservation wages are strongly shaped by reference points, supporting a causal interpretation of the reference wage hypothesis. In contrast, Clark, Frijters, and Shields (2008) and Godechot and Senik (2015) emphasize observational evidence that individuals’ wage expectations and satisfaction are influenced by social comparisons. Noy and Sin (2021), while also relying on observational data, offer evidence suggesting a causal effect of income comparisons—particularly with peers of the same age or gender—on subjective well-being.}

Our heterogeneity analysis (see Appendix C) further shows that even in occupations traditionally dominated by women, female job seekers tend to report lower desired wages than men. As discussed in Hara (2018), this gender gap cannot be fully explained by differences in human capital in Japan. Some prior studies have attributed such gaps to gender differences in competitiveness or bargaining behavior, which are known to vary across contexts and task types (Croson and Gneezy, 2009).\footnote{These behavioral tendencies are known to be context-sensitive, becoming more pronounced in mixed-gender settings (Booth and Nolen, 2012), and in tasks traditionally associated with men, where competitiveness and self-evaluation pressures are stronger (Günther et al., 2010; Große and Riener, 2010; Shurchkov, 2012; Exley and Kessler, 2022; Flory, Leibbrandt, and List, 2015). Some studies also suggest that the gender composition of the surrounding social structure influences such behaviors (Gneezy, Leonard, and List, 2009).} However, our finding that no gender gap is observed in female-typed occupations suggests that women are not simply lowering their desired wages because they are applying for male-dominated jobs or facing male competitors. 

Finally, our results align with Roussille (2024), who finds little evidence that preference differences explain the ask salary gap and instead emphasizes the role of information asymmetry. In her natural experiment, providing U.S. job seekers with information about average wages for similar workers eliminated the gender gap, suggesting that women were not inherently less ambitious but lacked reliable reference points. While her study focuses on occupation- and experience-based references, our findings suggest that Japanese job seekers may rely more on demographic cues such as age and gender. This difference may reflect institutional contexts, labor market fluidity, or access to information. Notably, our sample includes job seekers—who may compare themselves to broad market patterns—whereas Roussille examines incumbent workers, who tend to benchmark against their current positions (Fehr, Goette, and Zehnder 2009). These insights raise the question of whether providing transparent, gender-neutral reference information could help reduce Japan’s gender gap in desired wages—an avenue worth exploring in future research and policy design.

Taken together, these findings suggest that the gender gap in desired wages cannot be fully explained by differences in human capital, job preferences, or competitiveness. Instead, it reflects a combination of life-cycle expectations, compensating differentials, gendered social comparisons, and informational constraints. To address this gap, policies aimed solely at firm-side interventions may be insufficient. Promoting more equitable distributions of household responsibilities and providing clearer, gender-neutral reference information to job seekers could play a vital role in reducing gender-based disparities in wage expectations.

\section{Conclusion}\label{sec-conclusion}

This study investigates the salary disparity between men and women in
Japan using administrative data. Unlike previous research, our analysis
benefits from data that include various occupations that are not
necessarily high-skilled. We initially estimate the difference in
average desired wages to gauge the full extent of the gap, revealing
a notable 17.6 percentage point disparity. Given that the gender gap in
the average actual wage is 24.8 percentage points, this substantial
value indicates that the desired wage gap contributes significantly to Japan's
overall wage gap (although our desired wage gap result includes the unemployed
population).

To further understand these differences, a decomposition analysis using
the generalized KOB-Duncan method is conducted using information on the
three variables available, namely, age, desired region to work, and
desired occupation. The analysis reveals that a portion of the desired wage gap
could be explained by differences in the distribution of each variable
between men and women, i.e., -1.2 percentage points by age, 0.5
percentage points by desired region, 3.5 percentage points by desired
occupation, and 14.8 percentage points by other unexplained factors.
Among the differences in the distribution of the available information,
the primary influence of the desired occupation is consistent with the
general discussion of the wage gap between men and women (Blau and Kahn
2017).

Finally, we conduct a heterogeneity analysis to estimate the desired wage gap
within the three-factor combination groups to better understand the
impact of the unobservable factor, which is the largest factor in the
factor decomposition of the desired wage gap. Specifically, when we test 150
group combinations for the three variables, we find that the difference
widens in the middle-aged group. This result suggests that factors such
as compensating for the wage differential and gap in actual experience
from intrahousehold specialization, or/and differences in reference
wages may amplify the desired wage gap during this particular life stage.
Furthermore, we observe that women's desired wages are also low among
those who want to work in traditionally female-dominated occupations.

Notably, our study has several limitations. First, it is constrained by
the availability of only three pieces of information on workers. Future
analyses could benefit from incorporating more detailed information on
individuals to enhance the depth and accuracy of the study. Second, a
valuable avenue for improvement involves comparing the findings with
actual wage differentials derived from representative data on workers.
Such a comparison would provide a valuable benchmark, enabling a more
comprehensive understanding of the relationship between gaps in wage preferences and actual wages in the labor market.

\section*{Acknowledgments}
The authors are grateful to Hiroaki Miyamoto, Fumio Ohtake, Hideo Owan, and Masaru Sasaki for their insightful and constructive comments. 
We also thank all the participants of the JEA Spring Meeting in 2024.
We are grateful to the Ministry of Health, Labour and Welfare of Japan for granting access to the data used in this study.
An AI tool was used solely to assist with language editing; all interpretations, conclusions, and any remaining errors are the sole responsibility of the authors.

\section*{Funding}
This work was supported by JSPS KAKENHI [Grant Numbers 18K01661, 22H00854, 23K01418, and 25K05137].
The funders had no role in the study design, data collection and analysis,
decision to publish, or preparation of the manuscript.

\section*{Data availability}
Access to the data was granted by the Ministry of Health, Labour and Welfare of Japan upon approval of our research proposal.
The dataset cannot be distributed to third parties. 
Researchers who require access to the data should directly contact the Ministry of Health, Labour and Welfare of Japan.
To ensure reproducibility, we are happy to provide additional information, including analysis programs.

\section*{Conflict of interest}
The authors declare no conflicts of interest.

\newpage

\section*{Appendices}\label{appendicesa}
\addcontentsline{toc}{section}{Appendices}

\subsection*{Appendix A: Summary
Statistics}\label{appendix-a-summary-statistics}
\addcontentsline{toc}{subsection}{Appendix A: Summary Statistics}

Table~\ref{tbl-summary} is here

In the table, ``Desired Wage'' is the respondents' desired wage (unit is
1000 yen). ``Occupation'' is the average offered wage for the desired
occupation, while ``Unknown'' is a sample in which the respondent does
not have a desired occupation. ``Region'' represents the average offered
wage for the prefecture. The numbers in the table indicate the median
(bottom 25\%, top 25\%).

\subsection*{Appendix B: Five-year trend of desired wages}\label{appendix-b-five-year-trend-of-ask-salaries}
\addcontentsline{toc}{subsection}{Appendix B: Five-year trend of desired wages}

This appendix discusses the transition, specifically examining desired wages and related variables for the five years leading up to the 2019
data utilized in the primary analysis.

\begin{figure}[H]

\centering{

\includegraphics{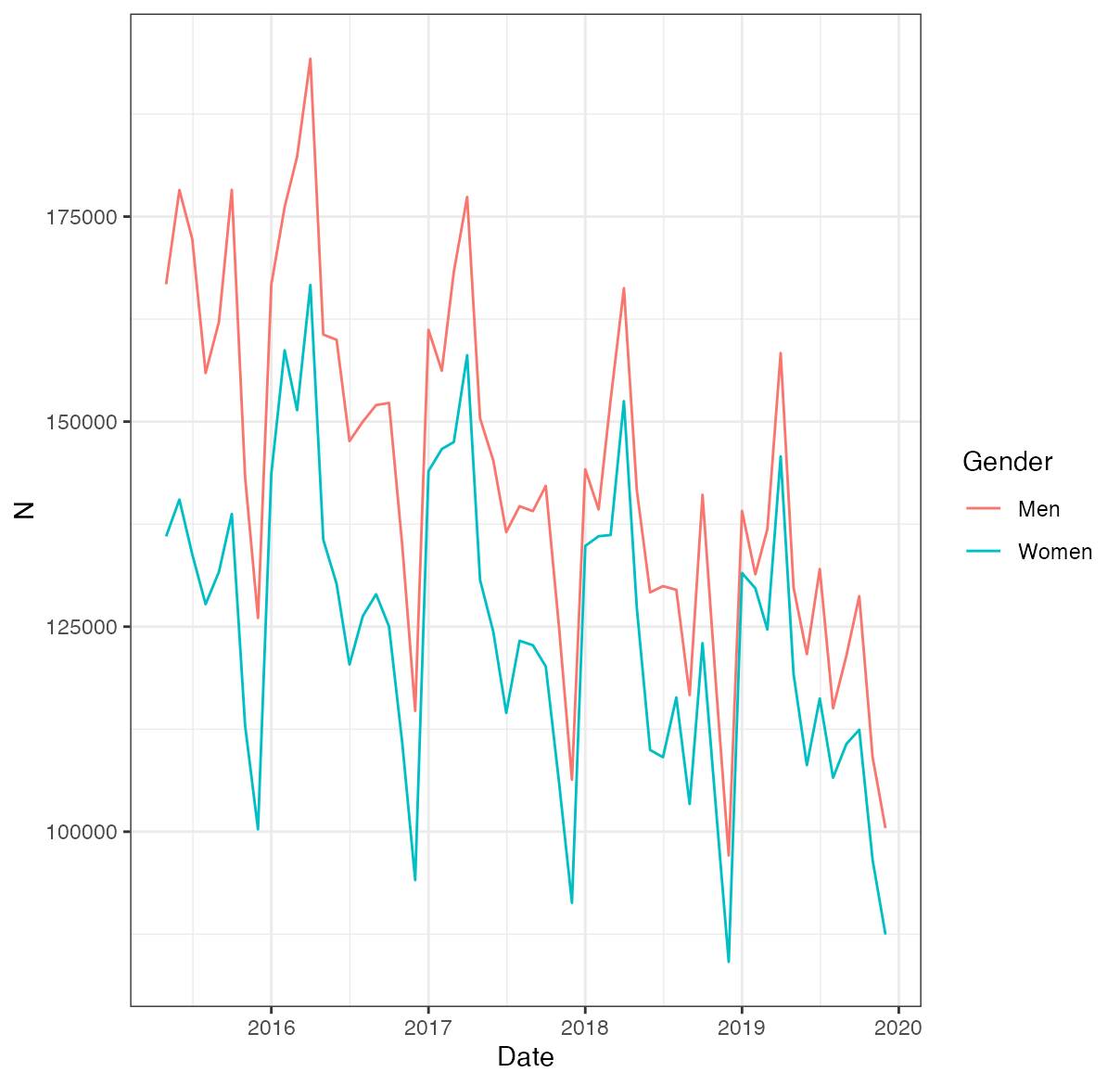}

}

\caption{\label{fig-Figure-n-seeker-5year}Number of job seekers in five
years}

\end{figure}

\begin{figure}[H]

\centering{

\includegraphics{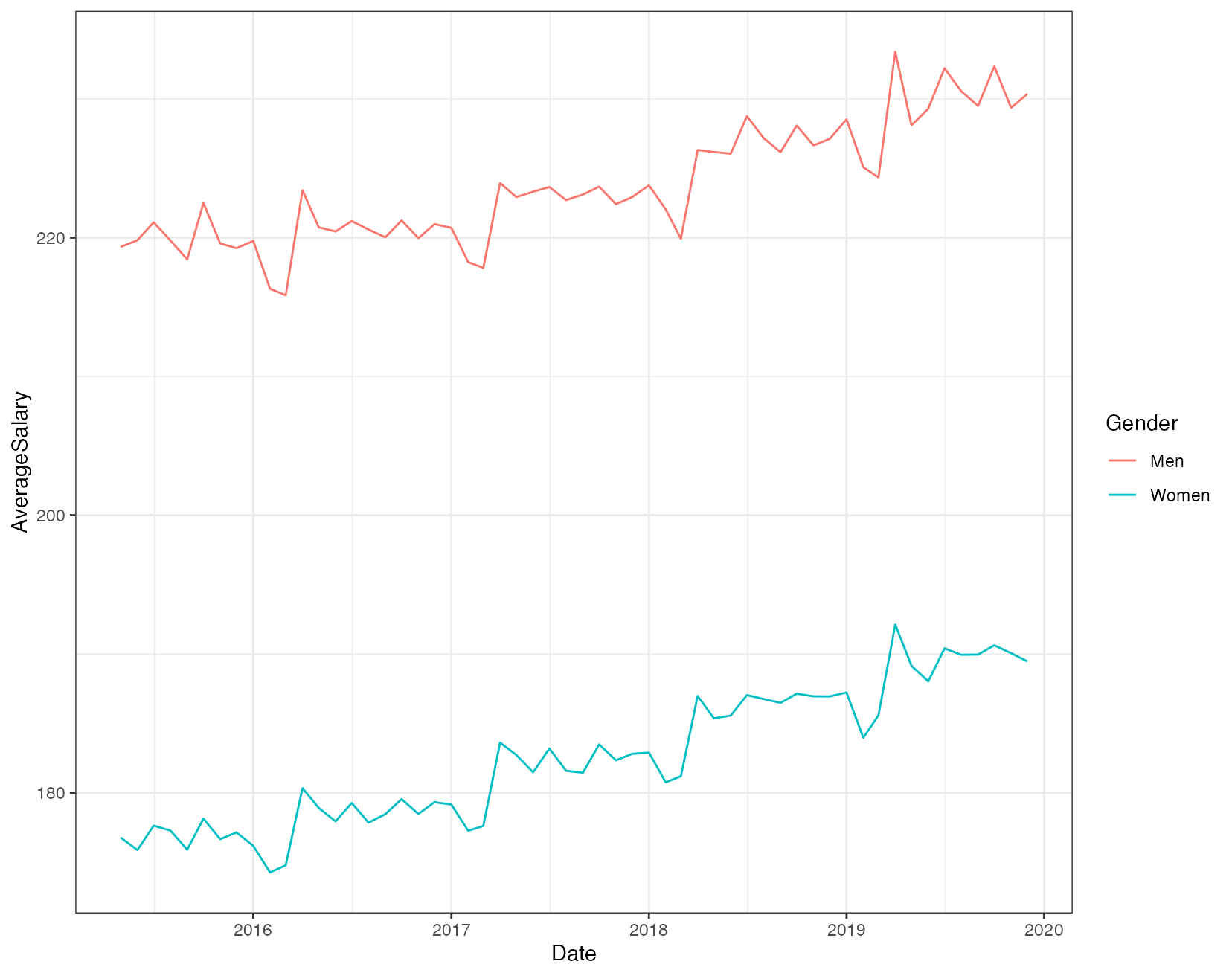}

}

\caption{\label{fig-Figure-askgap-5year}Trends of desired wages across a five-year
span}

\end{figure}

Figure~\ref{fig-Figure-n-seeker-5year} and
Figure~\ref{fig-Figure-askgap-5year} are around here.

First, we look at the number of new job seekers.
Figure~\ref{fig-Figure-n-seeker-5year} shows the number of new job
seekers by gender, with the vertical axis representing the number of new
job seekers, and the horizontal axis represents the corresponding year
and month. There is an overall downward trend in the number of job
seekers. Second, seasonality can be observed. For example, a spike is
observed in March for both men and women. This may be because the
Japanese fiscal year runs from April to March, which may affect the desired wages of those looking for a job upon retirement.

Next, let us look at the trends in desired wages, which are of interest
to us. In Figure~\ref{fig-Figure-askgap-5year}, the vertical axis is the
average monthly desired salaries, and the horizontal axis is the
corresponding year and month. Looking at
Figure~\ref{fig-Figure-askgap-5year}, we observe an overall increasing
trend. Furthermore, we observe seasonality similar to that seen in
Figure~\ref{fig-Figure-n-seeker-5year}, which may be due to the
demographics of new job seekers, as previously discussed.

\begin{figure}[H]

\centering{

\includegraphics{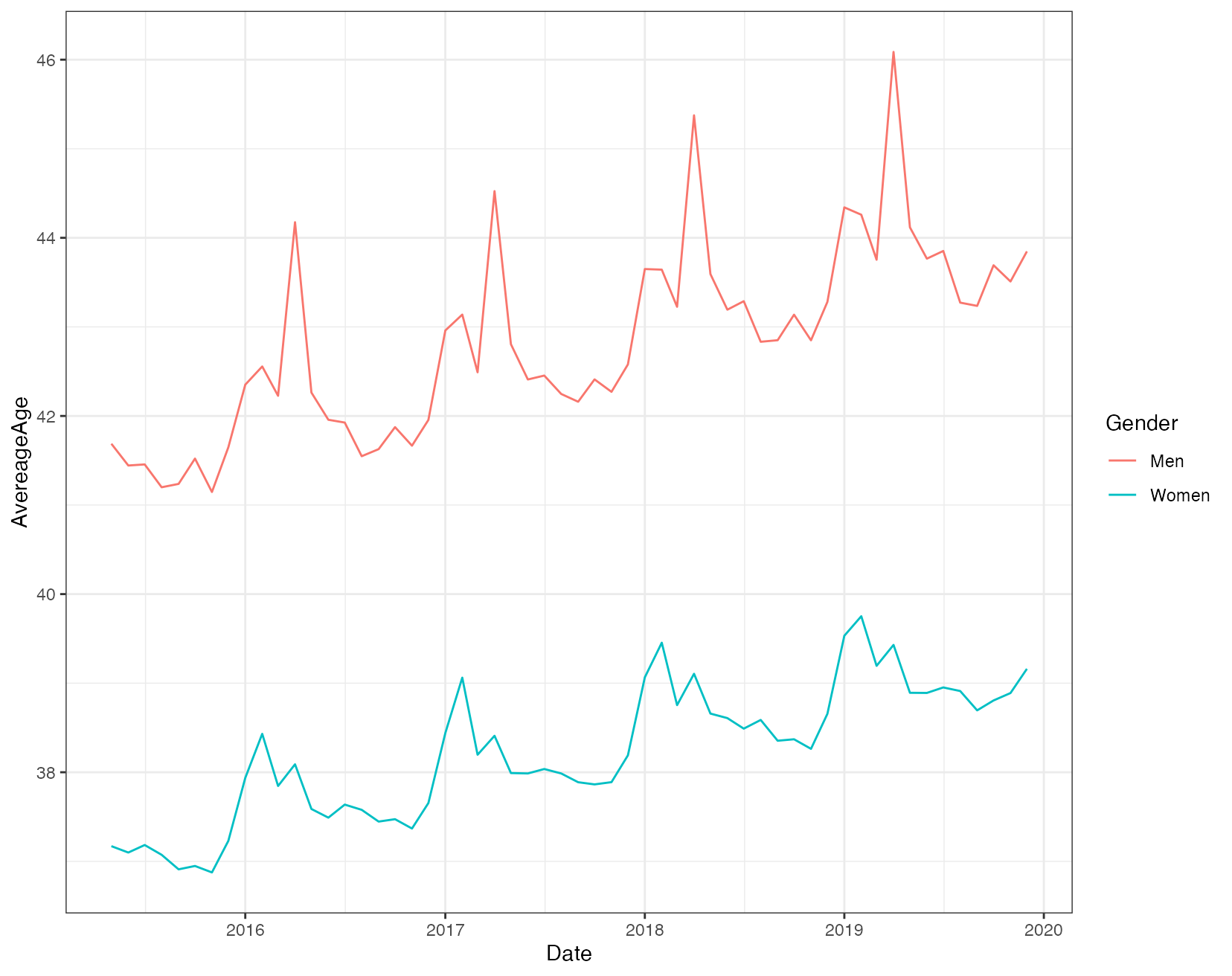}

}

\caption{\label{fig-Figure-age-5year}Trends of job seekers' age a five-year
span}

\end{figure}

To understand the factors behind the change in desired salaries, we attempt
to interpret it on the basis of the available data. First, we examine
the age of job seekers. Figure~\ref{fig-Figure-age-5year} illustrates
the average age of job seekers by gender on the vertical axis against
the corresponding year on the horizontal axis. It is evident from
Figure~\ref{fig-Figure-age-5year} that, overall, the average age of job
seekers is increasing for both men and women. This trend is likely
attributable to the aging population.

Figure~\ref{fig-Figure-age-5year} is around here.

\begin{figure}[H]

\centering{

\includegraphics{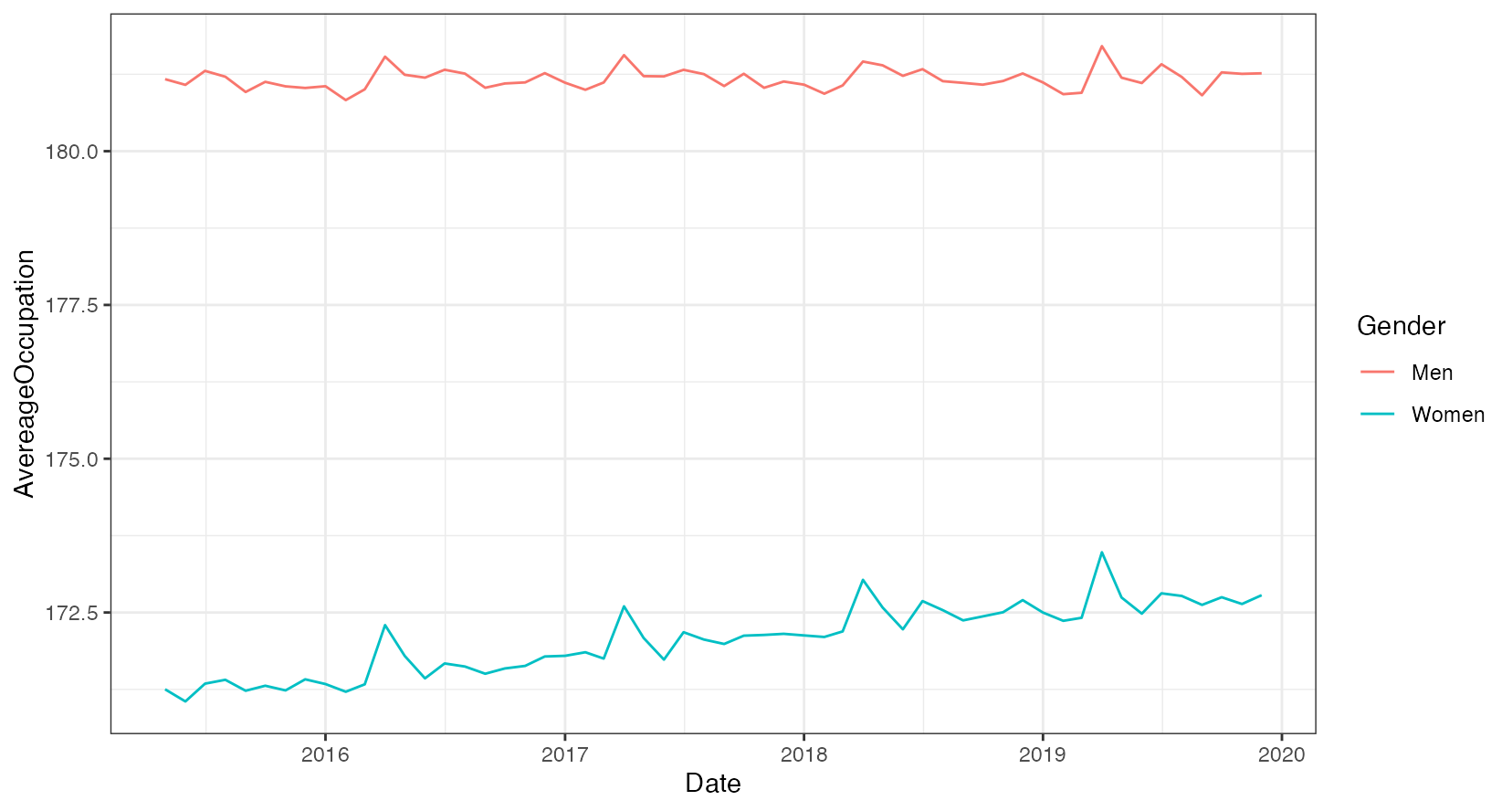}

}

\caption{\label{fig-Figure-occupation-5year}Trends of job seekers'
desired occupation across a five-year span}

\end{figure}

Second, we examine the changes in desired occupations.
Figure~\ref{fig-Figure-occupation-5year} displays the average offered
wage for job seekers' desired occupations by gender on the vertical
axis, with the corresponding year and month on the horizontal axis. The
average offered wage presented here is calculated solely from data drawn
from the initial dataset (May 2015). While the offered wages for men's
desired occupations have remained relatively stagnant, there has been a
steady increase in those for women. Notably, women are shown to be
gradually transitioning to jobs with higher wage offers.

Figure~\ref{fig-Figure-occupation-5year} is around here.

\begin{figure}[H]

\centering{

\includegraphics{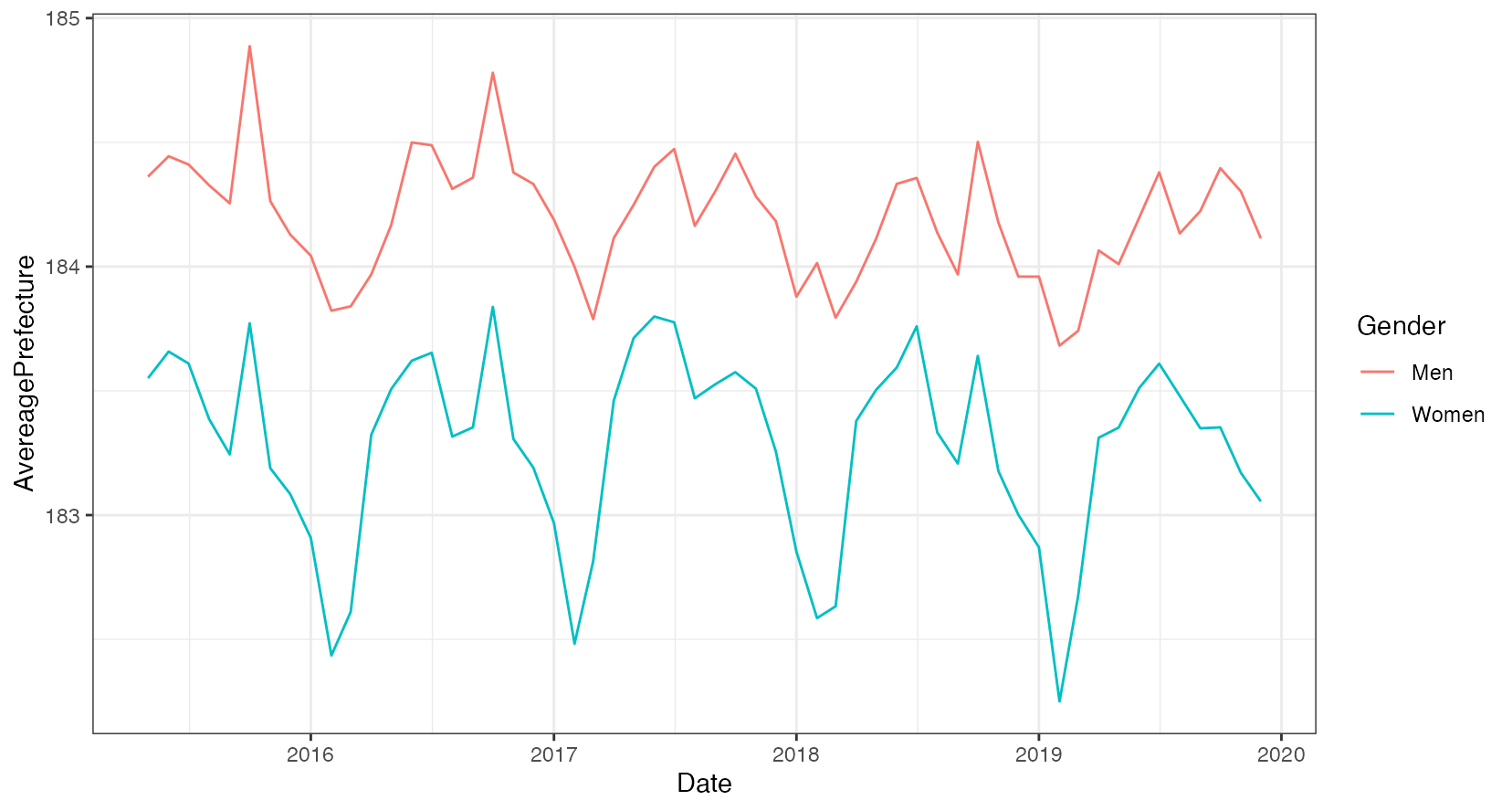}

}

\caption{\label{fig-Figure-region-5year}Trends of job seekers' desired
region across a five-year span}

\end{figure}

Finally, let us look at regional trends.
Figure~\ref{fig-Figure-region-5year} shows the average offered wage in
the job seeker's desired region to work (prefecture) by gender on the
vertical axis and the corresponding year and month on the horizontal
axis. Here, the average offered wage is calculated using only data from
the initial period (May 2015). Although seasonality can be observed for
the desired region (at a minor magnitude, however), there is no
significant trend in the overall change.

Figure~\ref{fig-Figure-region-5year} is around here.

Taken together with the three characteristics mentioned above, the shift
in desired occupations can be seen as a factor that influences the
overall trend. Regarding the wages offered for these desired jobs, there
has been a decreasing trend present over the past five years. However,
the primary outcome from the latest 2019 data indicates that a
discrepancy exists in regard to desired wages, which suggests that
desired occupations continue to have a certain impact.

\subsection*{Appendix C: Subgroup analysis for pink-collar
jobs}\label{appendix-c-subgroup-analysis-for-pink-collar-jobs}
\addcontentsline{toc}{subsection}{Appendix C: Subgroup analysis for
pink-collar jobs}

The primary findings in Section~\ref{sec-results} indicate that across
the entire sample, women's desired salaries are typically lower than men's.
However, do these findings remain consistent for ``pink-collar'\,' jobs,
which are traditionally occupied by women? As discussed in
Section~\ref{sec-hypotheses}, previous research has suggested that women
tend to be less competitive and less inclined to negotiate salaries than
men, but this inclination is understood to be influenced by the specific
context or environment. It has been observed that the gender disparity
vanishes when the competitor, the employer or supervisor with whom one
is negotiating is a woman. The analysis in this appendix investigates
gender disparities in desired wages among individuals seeking pink-collar
jobs in an environment where these women can be more proactive in
competition and wage negotiations.

Here, we focus on the subgroup consisting of childcare workers, nurses,
and caregivers among the pink-collar jobs. The reasons for focusing on
these three occupations are as follows. First, these occupations
traditionally employ a large percentage of women. Second, these three
occupations are in high demand and have a large effect on the labor
market. Third, these occupations are professional positions that require
qualifications, which is expected to make them relatively unaffected by
career intervals such as childbearing and to make it easier for them to
understand their market value.

The analysis follows the decomposition procedure described in
Section~\ref{sec-estimation_strategy} within the subgroup consisting of
the three occupational subgroups. The point estimator (and standard
error) is 0.088 (0.001) for the subgroup, suggesting that the women's
desired wages are still lower than those of men even in these
occupations.

\subsection*{Appendix D: Possible explanations of desired wage gap from existing literature}\label{appendix-d-possible-explanations-of-ask gap-from-existing-literature}
\addcontentsline{toc}{subsection}{Appendix D: Possible explanations of desired wage gap from existing literature}

There is a substantial body of research on the gender wage gap, which is
extensively summarized in Blau and Kahn (2017), among others. Various
factors have been identified to explain differences in labor outcomes
between men and women. In this Appendix, we organize and present the
existing findings, focusing on disparities in occupation, age, and
region between men and women that can affect the desired wage gap. Additionally,
we explore existing findings regarding gender differences within groups
with similar characteristics.

\subsubsection*{Occupation}\label{occupation}

First, it has been suggested that there is a difference in the
distribution of occupations among male and female workers. According to
Blau and Kahn (2017), 17\% and 32\% of the gender wage gap in the U.S. can
be accounted for by industry and occupation, respectively. England,
Levine, and Mishel (2020) offered an overview of gender disparity in the
U.S., noting a decline in occupational segregation from 1970 to 2015;
however, such segregation remains significant, with a Duncan index of
0.42. Some studies have interpreted job segregation as stemming from
differences in domestic responsibilities, such as child-rearing. Goldin
(2014), who explored the persistent gender wage gap in the U.S., found
that the cost of adjusting work hours, both in terms of length and
timing, varies across occupations. She regarded this as a crucial factor
that contributes to the gender wage gap between different occupations,
known as the compensated wage differentials. Kleven, Landais, and
Søgaard (2019) analyzed the ``motherhood dip'' phenomenon in Denmark,
revealing that occupational downgrading tends to occur when individuals
become parents, with many transitioning to the public sector, which is
often referred to as the ``mommy track.'' Given such occupational
segregation among workers, there is likely to be segregation in terms of
desired occupations among job seekers.

\subsubsection*{Age}\label{age}

Second, it has been proposed that there are differences in the age
distribution of workers and job seekers between men and women. A widely
acknowledged pattern is that women tend to exit the labor market during
middle age. Earlier work by Killingsworth and Heckman (1986), which
examined labor participation among men and women across ages and
cohorts, revealed that gender disparities typically arise in the middle
age bracket. According to Goldin and Mitchell (2017) and an OECD report (2018),
labor market exit among middle-aged workers has diminished in the United
States and other developed nations, with labor participation rates
leveling off. However, within the OECD, only Japan and South Korea
continue to exhibit an M-shaped curve. Similar findings have been
observed in numerous studies in Japan, indicating that this difference
has not been fully eradicated in the latest market (OECD 2019).

\subsubsection*{Region}\label{region}

Third, it is expected that differences in the regional distribution of
employment and job-seeking activities exist between men and women. For
instance, if various regions exhibit distinct norms and industries, then
it is likely that the employment rates and numbers of job seekers will
vary between genders. If these regional characteristics are correlated
with labor productivity and wages, they could contribute to the gender
gap in desired wages. Studies such as Fortin (2005), which was conducted
among OECD countries, and Fernández, Fogli, and Olivetti (2004) and
Boelmann, Raute, and Schonberg (2021), which were conducted within the
same country (U.S. and Germany, respectively), have shown the relationship
between norms and labor outcomes. However, in contrast, Blau and Kahn
(2017) focused on the United States, and suggested that gender
differences in regional distribution have limited explanatory power in
the U.S. gender wage gap.

\subsubsection*{Other factors}\label{other-factors}

While we have outlined three hypotheses that can be examined using the
available data, importantly, these hypotheses are interconnected. This
interconnection may contribute to the gender differences observed among
groups with similar characteristics. For example, the differences in
family responsibilities already mentioned can affect differences within
the same variable (i.e., occupation, age, and region) that we cannot
observe. Bertrand, Goldin, and Katz (2010) analyzed differences in
career paths between men and women for MBA holders in the U.S. at the
top of the wage distribution. They showed that there is almost no gender
wage gap at the beginning of MBA holders' careers but that the gender
gap increases with age. Studies have also highlighted women's career
interruptions and working time adjustment. Daniel, Lacuesta, and
Rodr\'{\i}guez-Planas (2013) showed that working time adjustment explains
about two-thirds of the motherhood dip in Spain. Cooke et al. (2009)
analyzed family migration, women's labor income, and the wage gap
between married couples in the U.S. and the U.K. and reported that
migration tends to increase household income at the cost of the wife's
income, especially in the U.S.; this suggests that migration is the
cause of the gender gap within the region. These effects could affect
the gap in desired wages in groups with similar characteristics.

Second, differences in preferences and attitudes may also play a
significant role in contributing to unobservable gender differences. It
has been observed that men and women exhibit disparities in risk
preferences, competitiveness, attitudes toward bargaining, and social
preferences (Croson and Gneezy 2009; Marianne 2011)\footnote{On average,
  women tend to shy away from competition and may be less inclined or
  less skilled in negotiating than men (Babcock and Laschever 2003);
  numerous studies have confirmed these trends (Leibbrandt and List
  2015; Dittrich, Knabe, and Leipold 2014; Card, Cardoso, and Kline
  2016).}. For instance, Niederle and Vesterlund (2007) conducted
laboratory experiments in the U.S. to investigate whether and why women
tend to be more hesitant to engage in competition than men. They
reported that women display lower levels of competitiveness and that
gender differences in performance and risk preferences do not account
for this, with male overconfidence emerging as an explanatory factor. In
fact, Cortés et al. (2023) reported that risk preference and
overconfidence play nonnegligible roles in explaining gender differences
in job search behavior through the formation of reservation wages. Card,
Cardoso, and Kline (2016) analyzed the gender wage gap in Portugal in
terms of firm sorting (between-firm gap) and bargaining (within-firm
gap) and concluded that sorting (between-firm gap) explains
approximately 15\%-20\% of the gender wage gap, whereas bargaining
(within-firm gap) is less influential but still explains approximately
6\% of the gender gap, depending on the specification. These variations
in preferences and attitudes may contribute to differences in desired wages.

Third, in line with the second factor of preference and attitudes,
disparities in beliefs between men and women may also influence the desired wage
gap. For instance, individuals may hold differing beliefs regarding
their own abilities and evaluation (Exley and Kessler 2022), or they may
adjust their wage expectations on the basis of the earnings of others as
reference points (Clark, Frijters, and Shields 2008; Clark and
d'Ambrosio 2015). If individuals reference the labor market conditions
or the wages of workers with similar attributes, then the wage
progression and promotions experienced by men and women currently in the
workforce, although not directly tied to the job seeker's salaries, may
serve as benchmarks for desired salaries (Fehr, Goette, and Zehnder 2009;
Godechot and Senik 2015; Noy and Sin 2021). Loprest (1992) examined
differences in wage growth and the returns to job mobility between men
and women in the U.S. They found that the wage growth among those who
remain at their jobs is relatively similar between genders or even
slightly greater for women, while male job changers experience more than
twice the wage growth when compared with that of female job changers.
Blau and Kahn (2017) noted that 14\% of the gender wage gap in the U.S.
can be attributed to differences in years of experience. The importance
of gender differences in the probability of promotion to management
positions is also emphasized in explaining the gender wage gap in Japan,
which is our country of interest, where long working hours and other
factors that are contrary to work-life balance are required to be
promoted to management positions (Yamaguchi 2016).

Fourth, policies can affect wage differentials among workers in the same
group. Blau and Kahn (2013) examined the impact of family policies on
labor outcomes for men and women. They demonstrated that the absence of
such policies tend to decrease the likelihood of American women
transitioning to lower-income occupations while also affecting
differences in age distribution. Additionally, although not specifically
targeting gender disparities, minimum wage policies can also affect the
gender wage gap (Blau and Kahn 1997). Typically, women are
disproportionately represented among minimum wage workers. D. Kawaguchi
and Mori (2009) discussed the characteristics of minimum wage workers in
Japan, noting that approximately 70\% of them are women. They
highlighted that these women are more likely to be employed in retail,
wholesale, and restaurant industries. Thus, improvements in the minimum
wage level could potentially impact the gender wage gap within the same
region.

Finally, monopsony power within the relevant market can also contribute
to gender disparities within regions and firms (Manning 2011). Webber
(2016) examined the gender wage gap in the U.S. through Manning's (2003)
monopsony framework. Women tend to exhibit lower firm-level labor supply
elasticity, enabling firms to exert greater monopsonic power. Their
research revealed that, upon decomposing labor supply elasticity,
women's greater search friction stems from lower wage offers.

\newpage

\section*{Tables}\label{tables}
\addcontentsline{toc}{section}{Tables}

\begin{table}[H]

\caption{\label{tbl-summary}Summary Statistics}

\centering{

\includegraphics{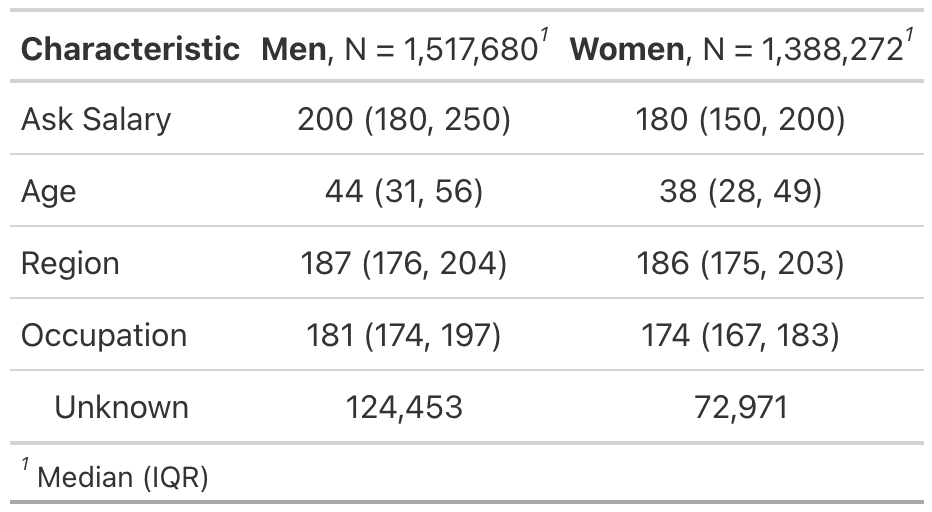}

}

\end{table}%

Note: Desired Wage is the respondents' desired wage (unit is 1000 yen).
Occupation is the average offered wage for the desired occupation, while
Unknown is a sample in which the respondent does not have a desired
occupation. Region represents the average offered wage for the
prefecture. The numbers in the table indicate the median (bottom 25\%,
top 25\%).

\newpage

\section*{Figures}\label{figures}
\addcontentsline{toc}{section}{Figures}

Notes: This figure shows the desired salaries by quartile of background
attributes presented above in a single figure. The labels on the upper
side line up with the quintiles of the regions running from left to
right. The labels on the right side line up with the occupational
quintiles running from top to bottom. The horizontal axis at the bottom
of the figure shows the age quintiles within each quintile of the
region. The vertical axis in each cell on the left side shows the
average desired salary in units of 1,000 yen. for each subgroup. The red and
blue dots denote males and females, respectively.

\newpage

Notes: This figure shows the ratio of women within each background
attribute to facilitate the interpretation of the generalized KOB-Duncan
method. The labels on the upper side line up with the quintiles of the
regions running from left to right. The labels on the right side line up
the occupational quintiles running from top to bottom. The horizontal
axis at the bottom of the figure shows the age quintiles within each
quintile of the region. Each plotted point in this figure shows the
ratio of women to men in the corresponding group; at a value of 0.5, the
ratio of men to women is the same.

\newpage

Note: This figure illustrates a box-and-whisker plot. This visual
representation showcases the difference between the logarithm of the
average wage for full-time workers in 2019, categorized by gender and
age group (as per our sample), and the logarithm of the desired wage for
individual job applicants within the same gender and age group.
Specifically, it illustrates the logarithmic value of each worker's desired
monthly salaries minus the logarithmic value of the average actual
monthly salaries.

\newpage

This figure illustrates a box-and-whisker plot. This visual representation showcases the difference between the logarithm of the average wage for full-time workers with zero years of tenure in 2019, categorized by gender and age group (as per our sample), and the logarithm of the desired wage for individual job applicants within the same gender and age group. The box represents the interquartile range (IQR), with the left and right edges indicating the 25th and 75th percentiles, respectively, and the vertical line inside the box denoting the median. Specifically, the figure illustrates the logarithmic value of each worker's desired monthly salary minus the logarithmic value of the average actual monthly salary.

\newpage

Notes: This figure shows the results of the raw desired wage gap and the desired
wage gap with all attributes aligned based on the data (i.e., age,
desired region to work, and desired occupation).

\newpage

Notes: This figure shows the results of the generalized KOB-Duncan
method. Specifically, we estimate Eq. (\ref{decomposition}) using the
generalized KOB-Duncan method. In the analysis, we utilize only the
subsamples that meet the positivity assumption criteria by excluding
subgroups with a male ratio exceeding 99.9\%. Decomposition analyses are
thus carried out by aligning the distribution of men with that of women.

\newpage

Notes: This figure shows the coefficients of the desired wage gap in salaries
between men and women for each subgroup presented in
Section~\ref{sec-data}. However, it is notable that we control for age,
desired region, and desired occupation in a nonparametric way.

\newpage

Notes: This figure shows the number of new job seekers by gender, with
the vertical axis representing the number of new job seekers and the
horizontal axis representing the corresponding year and month.

\newpage

Notes: In this figure, the vertical axis represents the average monthly
desired salaries, and the horizontal axis represents the corresponding year
and month.

\newpage

Notes: This figure illustrates the average age of job seekers by gender
on the vertical axis against the corresponding year on the horizontal
axis.

\newpage

Notes: This figure displays the average offered wage for the job
seekers' desired occupations by gender on the vertical axis, with the
corresponding year and month shown on the horizontal axis. The average
offered wage presented here is calculated solely using data from the
initial dataset (May 2015).

\newpage

Notes: This figure shows the average offered wage in the job seeker's
desired region to work (prefecture) by gender on the vertical axis and
the corresponding year and month on the horizontal axis. The average
offered wage here is calculated using only data from the initial period
(May 2015).

\newpage

\newpage

\section*{References}\label{references}
\addcontentsline{toc}{section}{References}

\phantomsection\label{refs}

\end{document}